\documentclass[runningheads]{llncs}

\usepackage{rotating}
\usepackage{graphics}
\usepackage{latexsym}
\usepackage[dvips]{epsfig}
\usepackage{amssymb}
\usepackage{graphicx}

\usepackage{url}

\def\boxit#1{\vbox{\hrule\hbox{\vrule\kern4pt
  \vbox{\kern1pt#1\kern1pt}
\kern2pt\vrule}\hrule}}

\def\qed{\rule{1.5mm}{3mm}}

\newcommand\nc{\newcommand}

\newtheorem{prob}[theorem]{Problem}

\nc{\crl}[2]{\begin{corollary}\label{crl:#1} #2 \end{corollary}}
\nc{\dfn}[2]{\begin{definition}\label{def:#1} #2 \end{definition}}
\nc{\lem}[2]{\begin{lemma}\label{lem:#1} #2 \end{lemma}}
\nc{\prp}[2]{\begin{proposition}\label{prp:#1} #2
\end{proposition}}
\nc{\thm}[2]{\begin{theorem}\label{thm:#1} #2\end{theorem}}
\nc{\fac}[2]{\begin{lemma}\label{fact:#1} #2 \end{lemma}}
\nc{\pro}[2]{\begin{prob}\label{#1} #2 \end{prob}}

\nc{\eqn}[2]{\begin{eqnarray}\label{eqn:#1} #2 \end{eqnarray}}
\nc{\fig}[4]{\begin{figure}[ht]
\begin{center}
\includegraphics[width=#2\textwidth]{#4}
\end{center}
\caption{#3}\label{fig:#1}
\end{figure}}

\nc{\tbl}[3]{\begin{table}[hbt] #3 \caption{#2} \label{tab:#1}
\end{table}}

\nc{\refc}[1]{Corollary~\ref{crl:#1}}
\nc{\refd}[1]{Definition~\ref{def:#1}}
\nc{\reff}[1]{Fig.~\ref{fig:#1}}
\nc{\refl}[1]{Lemma~\ref{lem:#1}}
\nc{\refp}[1]{Proposition~\ref{prp:#1}}
\nc{\reft}[1]{Theorem~\ref{thm:#1}} \nc{\refe}[1]{(\ref{eqn:#1})}
\nc{\reftb}[1]{Table~\ref{tab:#1}}

\nc{\reffc}[1]{Fact~\ref{fact:#1}}

\nc{\pf}[1]{ \noindent \emph{Proof.} #1
 \hfill \qed\par}

\begin{document}

%\mainmatter  % start of an individual contribution


\title{An Exact Algorithm for TSP in Degree-$3$ Graphs
via Circuit Procedure and Amortization on Connectivity Structure
\thanks{Supported in part by Grant
60903007 of NSFC, China.}
}

\author{Mingyu Xiao\inst{1} \and
Hiroshi Nagamochi\inst{2}}

 \institute{
 School of Computer Science and Engineering,
University of Electronic Science and Technology of China, China,
 \email{myxiao@gmail.com}
 \and
 Department of Applied Mathematics and Physics,
  Graduate School of Informatics, Kyoto University, Japan,
 \email{nag@amp.i.kyoto-u.ac.jp}}

\titlerunning{An Exact Algorithm for TSP in Degree-$3$ Graphs}

% the name(s) of the author(s) follo\omega(s) next
% \date{plain}

\authorrunning{}

% the affiliations are given next

%\institute{
%%Email: {\tt }%\footnote{Research partially supported by. }

\toctitle{TSP in Degree-$3$ Graphs} \tocauthor{}
\maketitle

\begin{abstract}
The paper presents an $O^*(1.2312^n)$-time and polynomial-space algorithm for the traveling salesman problem
 in an $n$-vertex graph with maximum degree $3$.
This improves the previous time bounds of
$O^*(1.251^n)$ by Iwama and Nakashima and $O^*(1.260^n)$ by Eppstein.
Our algorithm is a simple branch-and-search algorithm.
The only branch rule is designed on a cut-circuit structure of a graph induced by unprocessed edges.
To improve a time bound by a simple analysis on measure and conquer,
we introduce an amortization scheme over the cut-circuit structure
by defining the measure of an instance
to be the sum  of not only weights of vertices but also
weights of connected components of the induced graph.

\vspace{5mm}\noindent {\bf Key words.} Traveling Salesman Problem, Exact Exponential Algorithm, Graph Algorithm, Connectivity, Measure and Conquer
\end{abstract}

\section{Introduction}
The traveling salesman problem (TSP) is one of the most famous and intensively studied problems in computational mathematics. Many algorithmic methods have been investigated to beat this challenge of finding the shortest route visiting each member of a collection of $n$ locations and returning to the starting point.
 The first $O^*(2^n)$-time dynamic programming algorithm for TSP is back to early 1960s.
 However, in the last half of a century no one can break the barrier of $2$ in the base of the running time.
To make steps toward the long-standing and major open problem in exact exponential algorithms,
TSP in special classes of graphs, especially degree bounded graphs, have also been intensively studied.
Eppstein~\cite{Eppstein:TSP3} showed that TSP in degree-$3$ graphs (a graph with maximum degree $i$ is called a degree-$i$ graph) can be solved in $O^*(1.260^n)$ time and polynomial space, and TSP in degree-$4$ graphs can be solved in $O^*(1.890^n)$ time and polynomial space. Iwama and Nakashima~\cite{Iwama:TSP3} refined Eppstein's algorithm for degree-$3$ graphs and improved the result to $O^*(1.251^n)$ by showing that the worst case in Eppstein's algorithm will not always happen. Gebauer~\cite{Gebauer:TSP4} designed an $O^*(1.733^n)$-time exponential-space algorithm for TSP in degree-$4$ graphs,
which is improved to  $O^*(1.716^n)$ time and polynomial space by
Xiao and Nagamochi~\cite{XN:tsp4}.
Bjorklund \emph{et al.}~\cite{Bjorklund:TSPbounded} also showed TSP in degree bounded graph can be solved in $O^*((2-\varepsilon)^n)$ time, where $\varepsilon>0$ depends on the degree bound only. There is a Monte Carlo algorithm to decide a graph is Hamiltonian or not in $O^*(1.657^n)$ time~\cite{Bjorklund:Hamiltonicity}. For planar TSP and Euclidean TSP, there are sub-exponential algorithms based on small separators~\cite{Dorn:Planar}.
%Note that our improved algorithm for TSP in degree-4 graphs is obtained by
% successfully applying the measure and conquer method to TSP
%for the first time.
%We also showed that
%the analysis of the algorithm implies an $O^*(1.260^n)$-time algorithm
%for TSP in degree-3 graphs.

In this paper, we present an improved deterministic algorithm for TSP in degree-$3$ graphs, which runs in $O^*(2^{{{3}\over{10}}n})=O^*(1.2312^n)$ time and polynomial space. The algorithm is simple and contains only
one branch rule that is designed on a cut-circuit structure of a graph induced by unprocessed edges.
We will apply the measure and conquer method to analyze the running time. Note that our algorithm for TSP
in degree-4 graphs in~\cite{XN:tsp4} is obtained by successfully applying the measure and conquer method to TSP for the first time. However, direct application of measure and conquer to TSP in degree-3 graphs may only lead to an $O^*(1.260^n)$-time algorithm. To effectively analyze our algorithm, we use an amortization scheme over the cut-circuit structures by setting weights to both vertices and connected components of the induced graph.

%The paper is organized as follows.
%Section~\ref{sec:prelim} reviews basic notations
%and a polynomially solvable case.
%
%Section~\ref{sec:edge-conn} reviews the structure of 3-edge-connectivity
%of graphs, and define several important notions such as
%``circuits" and ``blocks."
%
%Section~\ref{sec:algo} introduces a procedure of processing
%edges in a circuit as a building block of  our algorithm,
%and gives an entire description of the algorithm.
%
%Section~\ref{sec:analysis} sets a weight function on vertices and
%components of unprocessed edges,
%and analyzes branch vectors of branching operations in our algorithm,
%
%Section~\ref{sec:conclude} makes some concluding remarks.

%\newpage

\section{Preliminaries}\label{sec:prelim}

In this paper, a graph means an undirected edge-weighted graph with maximum degree 3,
which possibly has multiple edges, but no self-loops.
%In such a graph, a set of edge-disjoint paths between two vertices is always a set of internally disjoint paths between them. {\tt [having no common internal vertices?]}
Let $G=(V,E)$ be a graph with an edge weight.
%We use $\mathrm{cost}(e)$ to denote the weight of an edge $e\in E$.
For a subset $V'\subseteq V$ of vertices and a subset $E'\subseteq E$ of edges,
the subgraphs induced by $V'$ and $E'$ are denoted by $G[V']$ and $G[E']$ respectively. We also use  $\mathrm{cost}(E')$ to denote the total weight of edges in $E'$.
For any graph $G'$, the sets of vertices
and edges in $G'$ are denoted as $V(G')$ and $E(G')$ respectively.
A graph consisting of a single vertex is called {\em trivial}.
A \emph{cycle} of length $l$ (also denoted as {\em $l$-cycle}) is a graph with $l$ vertices $v_i$ and $l$ edges $v_iv_{i+1}$ ($i\in \{1,2\dots,l\}$ and $v_{l+1}=v_1$). An edge $v_iv_j$ ($|i-j|\geq 2$) between two vertices in the cycle but different from the $l$ edges in it
is called a {\em chord} of the cycle.
Two vertices in a graph are
{\em $k$-edge-connected} if there are $k$-edge-disjoint paths between them. A graph is {\em $k$-edge-connected} if every pair of vertices in it are $k$-edge-connected.
We treat a trivial graph as a $k$-edge-connected graph for any $k\geq 1$.
A Hamiltonian cycle is a cycle through every vertex. Given a graph with an edge weight,
the \emph{traveling salesman problem} (TSP) is to find a Hamiltonian cycle of minimum total weight in the edges.

\subsection{Forced TSP}
%An instance of traveling salesman problem is a simple undirected graph with an edge cost.
In some branch-and-search algorithms for TSP, we may branch on an edge in the graph by including it to the solution
or excluding it from the solution. In this way, we need to maintain a set of edges that must be used in the solution.
We introduce the \emph{forced traveling salesman problem} as follows. An instance is a pair $(G,F)$ of an edge-weighted undirected graph
$G=(V,E)$ and
a subset $F\subseteq E$ of edges, called {\em forced edges}.
%Let $cost(e)$ denote the cost of an edge $e$ in $G$.
A Hamiltonian cycle of $G$ is called a {\em tour}
if it passes though all the forced edges in $F$. The objective of the problem is to compute a tour of minimum weight in
the given instance $(G,F)$. An instance is called {\em infeasible} if no tour exists. A vertex is called \emph{forced} if there is a forced edge incident on it. For convenience, we say that the \emph{sign} of an edge $e$ is 1 if $e$ is a forced edge and $0$ if $e$ is an unforced edge. We use $\mathrm{sign}(e)$ to denote the sign of $e$.

\subsection{$U$-graphs and $U$-components}
We consider an instance $(G,F)$. Let $U=E(G)-F$ denote the set of unforced edges.
A subgraph $H$ of $G$ is called a \emph{$U$-graph} if $H$ is a trivial graph or $H$ is
induced by a subset $U'\subseteq U$ of unforced edges (i.e., $H=G[U']$). A maximal connected $U$-graph is called a \emph{$U$-component}.
Note that each connected component in the graph $(V(G),U)$ is a $U$-component.

For a vertex subset $X$ (or a subgraph $X$) of $G$,
let $\mathrm{cut}(X)$ denote the set of edges in $E=F\cup U$
between $X$ and $V(G)-X$,
and denote  $\mathrm{cut}_F(X)=\mathrm{cut}(X)\cap F$
and $\mathrm{cut}_U(X)=\mathrm{cut}(X)\cap U$. Edge set $\mathrm{cut}(X)$ is also called a \emph{cut} of the graph.
We say that an edge is \emph{incident} on $X$ if the edge is in $\mathrm{cut}(X)$.
The {\em degree} $d(v)$ of a vertex $v$ is defined to be
$|\mathrm{cut}(\{v\})|$.
We also denote $d_F(v)=|\mathrm{cut}_F(\{v\})|$
and $d_U(v)=|\mathrm{cut}_U(\{v\})|$.
A $U$-graph $H$ is {\em $k$-pendent} if $|\mathrm{cut}_U(H)|=k$.
A  $U$-graph $H$ is called  {\em even} (resp., {\em odd}) if  $|\mathrm{cut}_F (H)|$ is even (resp., odd).
A $U$-component is 0-pendent.

In this paper, we will always keep every $U$-component 2-edge-connected.
%This property is important and many of our analyses are based on this assumption.
For simplicity, we may regard a maximal path of
forced edges between two vertices $u$ and $v$
as a single forced edge $uv$ in an instance $(G,F)$, since we can assume that
 $d_F(v)= 2$ always implies $d(v)=2$ for any vertex $v$.

\subsection{Circuits and blocks}
We consider a nontrivial $U$-component $H$ in an instance $(G,F)$.
A {\em circuit} ${\cal C}$ in $H$ is a maximal sequence
$e_1,e_2,\ldots ,e_p$ of edges $e_i=u_iv_i\in E(H)$ $(1\leq i\leq p)$
such that for each $e_{i}\in {\cal C}$ $(i\neq p)$,
the next edge $e_{i+1}\in {\cal C}$ is given
by a subgraph $B_i$ of $H$ such that
$\mathrm{cut}_U(B_i)=\{e_i,e_{i+1}\}$. See \reff{circuit} for an illustration.
%\marginpar{{\tt see PPT1}}
We say that each subgraph $B_i$ is a \emph{block} along ${\cal C}$ and vertices $v_i$ and $u_{i+1}$ are the {\em endpoints} of block $B_i$.
By the maximality of ${\cal C}$, we know that any two vertices in each block $B_i$ are 2-edge-connected in the induced subgraph $G[B_i]$. It is possible that a circuit in a 2-edge-connected graph $H$ may contain only one edge $e=u_1v_1$. For this case, vertices $u_1$ and $v_1$ are connected by three edge-disjoint paths in $H$ and the circuit is called \emph{trivial}, where the unique block is the $U$-component $H$.
Each nontrivial circuit contains at least two blocks, each of which is a 2-pendent subgraph of $H$.
In our algorithm, we will consider only nontrivial circuits ${\cal C}$.
When $H$ is 2-edge-connected, there are $p\geq 2$ different blocks along a nontrivial circuit ${\cal C}$, where  $u_1$ and $v_p$ are in the same block $B_p$ and $\mathrm{cut}_U(B_p)=\{e_p,e_{1}\}$.
%Each subgraph $B_i$ defined above is a block and vertices $v_i$ and $u_{i+1}$ are the {\em endpoints} of block $B_i$.
A block $B_i$ is called {\em trivial} if $|V(B_i)|=1$ and $d_F(v)=1$ for the only vertex $v$ in it ($v$ is of degree 3 in $G$).
A block $B_i$ is called {\em reducible} if $|V(B_i)|=1$ and $d_F(v)=0$ for the only vertex $v$ in it ($v$ is of degree 2 in $G$).
A block $B_i$ with $V(B_i)=\{v_i=u_{i+1}\}$ is either trivial or reducible in a  2-edge-connected graph.
%Note that if $B_i$ is nontrivial, the two endpoints $v_i$ and $u_{i+1}$ of it should be degree-3 vertices in $H$.

\vspace{-0mm}\fig{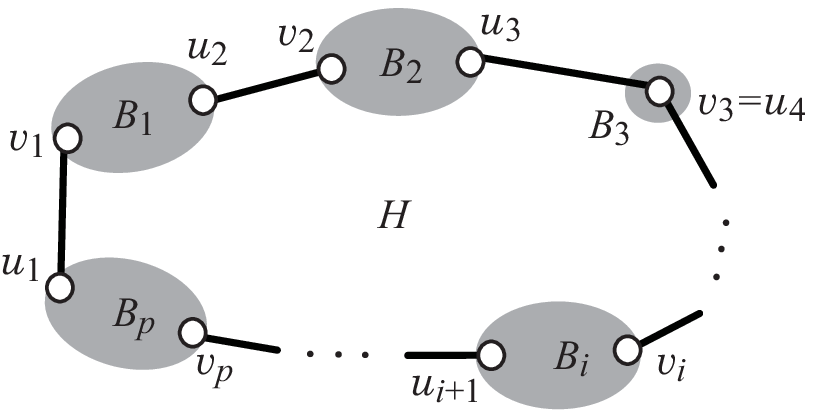}{0.4}{A circuit in a 2-edge-connected graph $H$}{circuit}\vspace{-0mm}
We state more properties on circuits and blocks.

\lem{circuits2}{In a degree-$3$ graph, let $H$ be a 2-edge-connected $U$-component and ${\cal C}$ be any circuit in it.
For each block $B_i$ of ${\cal C}$, $B_i$ is not trivial or reducible if and only if the two
endpoints $v_i$ and $u_{i+1}$ of it are two different vertices of degree $3$ in $H$.}

\lem{circuits1}{Each edge in a $2$-edge-connected $U$-component $H$ of a degree-$3$ graph is contained in exactly one circuit.
A partition of $E(H)$ into circuits can be obtained in linear time.}
\pf{
It is known that the set of all minimum cuts (a set of $k$ edges in a $k$-edge-connected multigraph
is called a minimum cut if the graph becomes disconnected by removing the $k$ edges)
can be represented by a cactus structure
(cf. \cite{NI:book}).
In particular, when the size of a minimum cut is two, the cactus structure
of minimum cuts can be
obtained in linear time by contracting each 3-edge-connected component
(a maximal set of vertices every two of which are 3-connected in the given graph)
into a single vertex, and for each cycle $C$ in the resulting graph,
a pair of any two edges in $C$ corresponds to a minimum cut in the original graph \cite{NI92}.
In a 2-edge-connected $U$-component $H$, (i) an edge $e\in E(H)$ forms a circuit ${\cal C}$ having only one block
if and only if $e$ is not in any minimum cut of $H$; and (ii)
A circuit ${\cal C}$ with at least two edges in $H$ corresponds to a cycle $C$ in the cactus structure.
Based on the cactus structure, we can obtain
a partition of edge sets into circuits in linear time.
% {\tt [NNN: some explanation is added] [XXX: Should we say something more here? We do not discuss how to partition the edge set yet.]}
}

\subsection{Branch-and-search algorithms}
Our algorithm is a branch-and-search algorithm: we search the solution by iteratively branching on the current
instance into several smaller instances until the current instance becomes trivial (or polynomially solvable).
In this paradigm, we will get a search tree. In each leaf of the search tree, we can solve the problem directly.
The size of the search tree is the exponential part of the running time of the search algorithm.
Let $\mu$ be a measure of the instance (for graph problems, the measure can be the number of vertices or edges in the graph and so on). Let $C(\mu)$ denote the maximum number of leaves in the search tree
generated by the algorithm for any instance with measure $\mu$.
We shall determine an upper bound on $C(\mu)$ by evaluating all the branches.
%We shall determine a factor $\alpha$ such that $C(\mu)=O^*(\alpha^{\mu})$.
When we branch on an instance $(G,F)$ with $k$ branches such that
the $i$-th branch decreases the measure $\mu$ of $(G,F)$ by
at least $a_i$, we obtain the following recurrence
$$C(\mu)\leq  C(\mu-a_1)+C(\mu-a_2)+\cdots +C(\mu-a_k).$$
Solving this recurrence, we get $C(\mu)=[\alpha(a_1, a_2, \ldots,
a_k)]^\mu$, where $\alpha(a_1, a_2, \ldots, a_k)$ is the largest
root of the function $f(x)=1-\sum_{i=1}^k x^{-a_i}$.
In this paper, we represent the above recurrence by a vector
$(a_1;a_2;\cdots ; a_k)$ of measure decreases, called a {\em branch vector}
(cf. \cite{Fomin:book}).
In particular, when $a_i=a_{i+1}=\cdots=a_j$ for some $i\leq j$,
it may be written as
$(a_1;a_2;\cdots a_{i-1};[a_i]_{j-i+1};a_{j+1}; \cdots ; a_k)$,
and a vector $([a]_k)$ is simply written as $[a]_k$.
When we compare two branch vectors $\mathbf{b}=(a_1;a_2)$ $(a_1\leq a_2)$
and $\mathbf{b}'=(a'_1;a'_2)$ such that ``$a_i\leq a'_i$ $(i=1,2)$" or ``$a'_1=a_1-\varepsilon$ and
$a'_2=a_2+\varepsilon$ for some $0\leq \varepsilon\leq a_2-a_1$,"
we only consider branch vector $\mathbf{b}$ in analysis,
since a solution $\alpha$ from  $\mathbf{b}$ is not
smaller than that from $\mathbf{b}'$ (cf. \cite{Fomin:book}).
We say that  $\mathbf{b}$ {\em covers} $\mathbf{b}'$ in this case.

\section{Reductions based on small cuts}
For some special cases, we can reduce the instance directly without branching.
Most of out reduction rules are based on the structures of small cuts in the graph.
In fact, we will deal with cuts of size $1,2,3$ and $4$.

\subsection{Sufficient conditions for infeasibility}
The {\em parity condition} on an instance is:
(i)  every $U$-component is even; and
(ii)  the number of odd blocks along every circuit is even.

\lem{infeasibility}{An instance $(G,F)$ is infeasible if
$G$ is not 2-edge-connected or it violates the parity condition.}
\pf{Since any tour is a 2-edge-connected spanning graph of $G$,
it cannot exist when $G$ is not 2-edge-connected.
Since any tour is an Eulerian graph,
it cannot exist in any instance with an odd $U$-component.
For a circuit which has an odd number of odd blocks,
we see that at least one odd block will be an odd $U$-component
in any way of including/deleting edges in the circuit.}

\subsection{Eliminable, reducible and parallel edges}
The unique unforced edge incident on a 1-pendent $U$-graph is \emph{eliminable}.
From parity condition (i), we can decide whether each eliminable edge need to be included to $F$ or deleted from the graph
just by depending on the parity of $|\mathrm{cut}_F(H)|$.

For any subgraph $H$ of $G$ with $|\mathrm{cut}(H)|=2$,
we call the unforced edges in $\mathrm{cut}(H)$ {\em reducible}.
From the connectivity condition and parity condition (i),
we see that all reducible edges need to be included to $F$.
In particular, any edge $uv$ incident to a vertex $v$ with $d(v)=2$
(or with a neighbor $v'$ with multiple edges
of $vv'\in F$ and $vv'\in U$) is reducible,
since $uv\in \mathrm{cut}(X)$ and $|\mathrm{cut}(X)|=2$
 for $X=\{v\}$ (or $X=\{v,v'\}$).

%When the graph has only two vertices, the problem can be solved easily. Else we assume the graph has more than two vertices. If there are some multiple unforced edges
%with the same endpoints, we can simply remove one with the heaviest cost preserving the optimality of the instance.

If there are multiple edges with the same endpoints $u$ and $v$, we can reduce the instance in the following way
 preserving the optimality:
 If the graph has only two vertices $u$ and $v$, solve the problem directly; else if there are forced edges between $u$ and $v$, the problem is infeasible; and otherwise remove all unforced edges between $u$ and $v$ except one with the smallest weight.

\subsection{Reductions based on 3-cuts and 4-cuts}
\lem{3subgraph}{Let $(G,F)$ be an instance where $G$ is a graph with maximum degree 3.
For any subgraph $X$ with $|\mathrm{cut}(X)|=3$, we can replace $X$ with a single vertex $x$ and update the
three edges incident on $x$ preserving the optimality of the instance.}

\pf{
Denote  $\mathrm{cut}(X)$ by $\{y_1x_1,y_2x_2,y_3x_3\}$ with $x_i\in V(X)$ and $y_i\in V-V(X)$.
We will replace $X$ and $\mathrm{cut}(X)$ with a single vertex $x$ and  three new edges $xy_1,xy_2$ and $xy_3$.
 Let $G'$ denote the new graph. We only need to decide the weights and signs of edges $xy_1,xy_2$ and $xy_3$ in $G'$ to satisfy the lemma.
Let $I_i$ ($i=1,2,3$) denote the problem of finding a path $P$ from $x_{i_1}$ to $x_{i_2}$ ($\{i,i_1,i_2\}=\{1,2,3\}$) of minimum total cost in $X$ that passes through all vertices and forced edges in $X$. We say that $I_i$ \emph{infeasible} if it has no such path.
We consider the three problems $I_i$ ($i=1,2,3$).
There are four possible cases.
Case 1. None of the three problems is feasible: We can see that the original instance $(G,F)$ is also infeasible. In $G'$, we let $\mathrm{sign}(xy_1)=\mathrm{sign}(xy_2)=\mathrm{sign}(xy_3)=1$.
 Since the trivial $U$-component $\{x\}$ is odd, the new instance is infeasible
by \refl{infeasibility}.
 Case 2. Only one of the three problems, say $I_{j_1}$ ($\{j_1,j_2,j_3\}=\{1,2,3\}$), is feasible: Let $S_{j_1}$ be an optimal solution to $I_{j_1}$. Then there is a solution $S$ to $(G,F)$  such that $S\cap E(X)=E(S_{j_1})$ (if $(G,F)$ is feasible). Therefore, in $G'$, we let $\mathrm{sign}(xy_{j_2})=\mathrm{sign}(xy_{j_3})=1$,
$\mathrm{sign}(xy_{j_1})=\mathrm{sign}(x_{j_1}y_{j_1})$,
$\mathrm{cost}(xy_{j_2})=\mathrm{cost}(x_{j_2}y_{j_2})+\mathrm{cost}(S_{j_1})$,
 $\mathrm{cost}(xy_{j_3})=\mathrm{cost}(x_{j_3}y_{j_3})$ and $\mathrm{cost}(xy_{j_1})=\mathrm{cost}(x_{j_1}y_{j_1})$.
Case 3. Exactly two of the three problems, say $I_{j_1}$ and $I_{j_2}$ ($\{j_1,j_2,j_3\}=\{1,2,3\}$), are feasible:
Let $S_{j_1}$ and $S_{j_2}$ be an optimal solution to $I_{j_1}$ and $I_{j_2}$ respectively.
Then there is a solution $S$ to $(G,F)$ such that either $S\cap E(X)=E(S_{j_1})$ or $S\cap E(X)=E(S_{j_2})$. Therefore,
in $G'$, we let $\mathrm{sign}(xy_{j_3})=1$, $\mathrm{sign}(xy_{j_1})=\mathrm{sign}(x_{j_1}y_{j_1})$, $\mathrm{sign}(xy_{j_2})=\mathrm{sign}(x_{j_2}y_{j_2})$, $\mathrm{cost}(xy_{j_3})=\mathrm{cost}(x_{j_3}y_{j_3})$,
$\mathrm{cost}(xy_{j_1})=\mathrm{cost}(x_{j_1}y_{j_1})+\mathrm{cost}(S_{j_2})$ and
$\mathrm{cost}(xy_{j_2})=\mathrm{cost}(x_{j_2}y_{j_2})+\mathrm{cost}(S_{j_1})$.
 Case 4. All of the three problems are feasible: Let $S_{1}$, $S_2$ and $S_{3}$ be an optimal solution to $I_{1}$, $I_2$ and $I_{3}$ respectively.
In $G'$, we let $\mathrm{sign}(xy_{i})=\mathrm{sign}(x_{i}y_{i})$ and
$\mathrm{cost}(xy_{i})= \mathrm{cost}(x_iy_{i})+{1\over 2}{\sum_{j=1}^3 \mathrm{cost}(S_{j}) }- \mathrm{cost}(S_{i})$ ($i=1,2,3$).
Straightforward computation can verify that with these setting $G'$ will preserve the optimality.
}
\bigskip

Similar to \refl{3subgraph}, we can simplify a subgraph $X$ with $|\mathrm{cut}(X)|=4$. However, there are
many cases needed to consider. In fact, in our algorithms, we only need to consider a special case.
%{\tt[XXX: The following parts are modified.]}

We consider a subgraph $X$ with $|\mathrm{cut}_F(X)|=4$ and $|\mathrm{cut}_U(X)|=0$.
We want to reduce $X$.
Denote $\mathrm{cut}(X)$ by $\{y_1x_1,y_2x_2,y_3x_3,y_4x_4\}$ with $x_i\in V(X)$ and $y_i\in V-V(X)$,
where $x_i\neq x_j$ $(1\leq i<j\leq 4)$.
We define $I_i$ ($i=1,2,3$) to be instances of the problem of finding two disjoint paths $P$ and $P'$ of minimum total cost in $X$ such that all vertices and forced edges in $X$  appear in exactly one of the two paths, and one of the two paths is from $x_i$ to $x_4$ and the other one is from $x_{j_1}$ to $x_{j_2}$ ($\{j_1,j_2\}=\{1,2,3\}-\{i\}$).
We say that $I_i$ \emph{infeasible} if it has no solution.

A subgraph $X$ is \emph{$4$-cut reducible} if $|\mathrm{cut}_F(X)|=4$, $|\mathrm{cut}_U(X)|=0$, and at least one of the three problems $I_1, I_2$ and $I_3$ defined above is infeasible.
We have the following lemma to reduce the $4$-cut reducible subgraph.

\lem{4subgraph}{
Let $(G,F)$ be an instance where $G$ is a graph with maximum degree 3.
A $4$-cut reducible subgraph $X$ can be replaced with
   one of the following subgraphs $X'$ with four vertices and
$|\mathrm{cut}_F(X')|=4$ so that the optimality of the instance is preserved:\\
{\rm (i)} four single vertices  $($i.e., there is no solution to this instance$)$;\\
{\rm (ii)}  a pair of forced edges; and \\
{\rm (iii)}  a 4-cycle with four unforced edges.
}
\pf{ We consider the three problems $I_i$ ($i=1,2,3$). Since at least one of them is infeasible, there are three possible cases.
Case 1. None of the three problems is feasible: We can see that the original instance $(G,F)$ is also infeasible.
Then we can replace $X$ with a graph containing only four vertices $\{x_1,x_2,x_3,x_4\}$ and no edge.
Now $x_1$ becomes a degree-1 vertex in the new graph and the new instance is infeasible.
Case 2. Only one of the three problems, say $I_{i_0}$ ($i_0\in\{1,2,3\}$), is feasible: Let $S_{i_0}$ be an
optimal solution to $I_{i_0}$. Then there is a solution $S$ to $(G,F)$ such that $S\cap E(X)=E(S_{i_0})$. Therefore, we can replace $X$ with a graph of four vertices $\{x_1,x_2,x_3,x_4\}$ and
 two edges $x_i x_4$ and $x_{j_1}x_{j_2}$ in $G$ preserving the optimality of the instance,
 where the costs of $x_{i_0} x_4$ and $x_{j_1}x_{j_2}$ are the costs of the two paths in $S_{i_0}$.
Note that in the new instance after the replacement, the four vertices $\{x_1,x_2,x_3,x_4\}$ become
degree-2 vertices and the two new edges $x_i x_4$ and $x_{j_1}x_{j_2}$ should be included into $F$.
Case 3. Two of the three problems, say $I_{i_1}$ and $I_{i_2}$ ($i_1,i_2\in\{1,2,3\}$), are feasible: Let $S_{i_1}$ and $S_{i_2}$ be  optimal solutions to $I_{i_1}$ and $I_{i_2}$ respectively. Then there is a solution $S$ to $(G,F)$ such that   $S\cap E(X)=E(S_{i_1})$ or $S\cap E(X)=E(S_{i_2})$. Therefore, we can replace $X$ with a $4$-cycle $x_{i_1} x_4 x_{i_2} x_j $ preserving the optimality of the instance, where $\{j\}=\{1,2,3\}-\{i_1,i_2\}$, the costs of $x_{i_1} x_4$ and $x_{i_2}x_{j}$ are the costs of the two paths in $S_{i_1}$, and the costs of $x_4 x_{i_2}$ and $x_{j}x_{i_1}$ are the costs of the two paths in $S_{i_2}$.
}
\bigskip

\lem{reducecritical}{
Let $X$ be an induced subgraph of a degree-$3$ graph $G$ such that
$X$ contains at most eight vertices of degree 3 in $G$.
Then $X$  is $4$-cut reducible if
 $|\mathrm{cut}_F(X)|=4$, $|\mathrm{cut}_U(X)|=0$,
and $X$ contains at most two unforced  vertices.
}
\pf{We only need to show that
at least one of the three problem instances $I_1, I_2$ and $I_3$ (defined before \refl{4subgraph}) is infeasible.
Assume that no two edges in $\mathrm{cut}_F(X)$ meet at a same vertex in $X$, since otherwise
only one of $I_1, I_2$ and $I_3$ is feasible.
Also assume that $X$ has no multiple edges or induced triangles, since otherwise
$X$ can be reduced to a smaller graph preserving its optimality.
Note that  $X$ contains an even number $k$ of degree 3 vertices in $G$.
Since $k=4$ (i.e., $|V(X)|=4$) implies the lemma, we consider the case of $k=6,8$.
When $k=6$, $X$ is either a 6-cycle with a chord or a graph obtained from
a 5-cycle with a chord by subdividing the chord with a new vertex.
In any case, we see that one of  $I_1, I_2$ and $I_3$ is infeasible.
Let $k=8$.
In this case, there are four vertices $u_i$ $(i=1,2,3,4)$ which are not incident to any of the four edges
in $\mathrm{cut}(B)=\mathrm{cut}_F(B)$, and two of them, say $u_1$ and $u_2$ are joined by
a forced edge $u_1u_2\in F$ by the assumption on the number of unforced vertices in $X$.
Then we see that there are three possible configurations for such an induced graph $X$ with no
induced triangles, and a straightforward inspection shows that none of them admits
a set of three feasible instances $I_1, I_2$ and $I_3$.
}
\bigskip

%{\rm (i)} $B$ is a $6$-cycle or extension of a $6$-cycle and has exactly one  forced chord; or
% {\rm (ii)} $B$ is obtained from a 2-pendent critical graph $B^*$ by deleting $\mathrm{cut}_U(B^*)$.

\refl{3subgraph} and \refl{4subgraph} imply a way of simplifying some local structures of an instance. However, it is not easy to find solutions to problems $I_i$ in the above two lemmas. In our algorithm, we only do this replacement for $X$ containing no more than 10 vertices and then the corresponding problems $I_i$ can be solved in constant time by a brute force search.

We define the operation of \emph{$3$/$4$-cut reduction}: If there a subgraph $X$ of $G$ with $|V(X)|\leq 10$ such that $|\mathrm{cut}(X)|=3$  or $X$ is $4$-cut reducible, then we simplify the graph by replacing $X$ with a graph according to \refl{3subgraph}  or \refl{4subgraph}. Note that a $3$/$4$-cut can be found in polynomial time if it exists and then this reduction operation can be implemented in polynomial time.

\subsection{A solvable case and reduced graphs}
A 3/4-cut reduction reduce the subgraph $X$ to a trivial graph except for the last case of \refl{4subgraph} where $X$ will become a 4-cycle.
Eppstein has identified a polynomially solvable case of forced TSP~\cite{Eppstein:TSP3}, which can deal with $U$-components of 4-cycles.

\lem{solvable}{{\rm \cite{Eppstein:TSP3}}
If every $U$-component is a component of a 4-cycle, then
a minimum cost tour of the instance can be found in polynomial time.
}

Based on this lemma, we do not need to deal with $U$-components of 4-cycles in our algorithms.

\bigskip
All above reduction rules can be applied in polynomial time. An instance $(G,F)$ is called a \emph{reduced} instance if
%{\small\tt [NNN: instances may be infeasible (testing feasiblity is part of the problem)]}
$G$ is 2-edge-connected, $(G,F)$ satisfies the parity condition, and
  has none of reducible edges, eliminable edges and multiple edges, and the 3/4-cut reduction cannot be applied on it anymore.
Note that a reduced instance has no triangle, otherwise 3-cut reduction would be applicable.
An instance is called {\em 2-edge-connected}
if
%it has no multiple unforced edges and
every $U$-component in it is 2-edge-connected.
The initial instance $(G,F=\emptyset)$ is assumed to be 2-edge-connected, otherwise
it is infeasible by \refl{infeasibility}.
In our algorithm, we will guarantee that the input instance is always 2-edge-connected,
and we branch on a reduced graph to search a solution.

%\lem{reducedinstance}{An instance is a $2$-edge-connected instance if and only if it has neither eliminable edges nor parallel unforced edges.}

%{\tt[XXX: I temporarily removed the definition of 2-edge-connected instances to avoid confusing it with the 2-edge-connected $G$ in \refl{infeasibility}.]}

%\section{The Algorithm}\label{sec:algo}
%Before presenting the whole algorithm, we introduce the basic operations used in it.
%In fact, except the basic reduction operations, our algorithm only contains one basic operation that is the \emph{circuit procedure}.

\section{The circuit procedure}
The \emph{circuit procedure} is one of the most important operations in our algorithm.
The procedure will determinate each edge in a circuit to be included into $F$ or to be deleted from the graph.
It will be widely used as the only branching operation in our algorithm.

\paragraph{Processing circuits:}
\emph{Determining} an unforced edge means either including it to $F$ or deleting it from the graph.
 When an edge is determined, the other edges in the same circuit containing this edge can also be determined directly by reducing eliminable edges. We call the series of procedures applied to all edges in a circuit together as a \emph{circuit procedure}.
Thus, in the circuit procedure, after we start to process a circuit ${\cal C}$ either by including an edge $e_1\in {\cal C}$ to $F$
 or by deleting $e_1$ from the graph, the next edge $e_{i+1}$ of $e_i$ becomes an eliminable edge and we continue to determine $e_{i+1}$
either by deleting it  from the graph if block $B_i$ is odd and $e_i=u_{i}v_i$
 is included to $F$ (or $B_i$ is even and $e_i$ is deleted);
or  by including it to $F$ otherwise.
Circuit procedure is a fundamental operation
 to build up our proposed algorithm. Note that a circuit procedure determines only the edges in the circuit.
During the procedure, some unforced edges outside the circuit may become reducible  and so on,
 but we do not determine them in this execution.

 \lem{circuitprocess}{
Let $H$ be a 2-edge-connected $U$-component in an instance $(G,F)$ and ${\cal C}$ be a circuit in $H$. Let $(G',F')$ be the resulting instance after applying circuit procedure on ${\cal C}$. Then\\
{\rm (i)} each block $B_i$ of ${\cal C}$ becomes a 2-edge-connected $U$-component in $(G',F')$; and \\
{\rm (ii)} any other $U$-component $H'$ than $H$ in $(G,F)$ remains unchanged in $(G',F')$.
}
\pf{Since $H$ is 2-edge-connected, we know that each block $B_i$ induces a 2-edge-connected subgraph
from $H$ according to the definition of circuits.
Hence  $B_i$ will be a 2-edge-connected $U$-component in $(G',F')$.
 Then we get (i). Since $H$ and $H'$ are vertex-disjoint and
only edges in $H$ are determined, we see that (ii) holds.
}\bigskip

We call a circuit \emph{reducible} if it contains at least one reducible edge.
We can apply the circuit procedure on a reducible circuit directly starting by including a reducible edge to $F$.
In our algorithm, we will deal with reducible edges
by processing a reducible circuit.
%\emph{Reducing a reducible circuit} means applying the circuit procedure on it starting by including a reducible edge in the circuit to $F$.
When the instance becomes a reduced instance, we may not be able to reduce the instance directly.
Then we search the solution by ``branching on a circuit."
\emph{Branching on a circuit ${\cal C}$ at edge $e\in {\cal C}$} means branching on the current instance to generate
 two instances by applying the circuit procedure to ${\cal C}$ after including $e$ to $F$ and deleting $e$ from the graph respectively.
%Note that we apply the circuit procedure in each branch automatically after the branch.
Branching on a circuit is the only branching operation used in our algorithm.
%s that may cause exponential running time.

%The simplest rule is to \emph{branch on an unforced edge} by including it to $F$ or deleting it from the graph.
%However, when an edge is determinated (by included to $F$ or deleted), some other edges, say the edges in the circuit containing the edge, can also be determinated by applying the reduction procedure.
%For the purpose of analysis, we may combine the processes of all edges in a circuit together to form the \emph{circuit procedure}.

%\lem{circuit}{Let ${\cal C}$ be a circuit
% in a 2-edge-connected instance $(G,F)$.
%The resulting instance $(G',F')$ by processing a circuit
%remains  2-edge-connected.
%Moreover all reducible edges along the circuit
%has been eliminated in $(G',F')$ without creating new reducible edges.
%}\bigskip

% \newpage

\section{A simple algorithm based on circuit procedures}\label{simple_alg}
We first introduce a simple algorithm for forced TSP to show the effectiveness of the circuit procedures. Improved algorithms are given in the next sections.

The simple algorithm contains only two steps: First reduce the instance until it becomes a reduced one;
 %or each $U$-component is trivial or a 4-cycle;
 and then select a $U$-component $H$ that is neither trivial nor a 4-cycle % (if it exists)
 and branch on a circuit ${\cal C}$ in $H$ such that at least one block along ${\cal C}$ is trivial.
Note that %we can simply assume that
there is always a circuit having a trivial block as long as the forced edge set $F$ is not empty.
% (the initial graph is connected).

Here we use a traditional method to analyze the simple algorithm. It is natural to consider how many edges can be added to $F$ in each operation of the algorithm. The size of $F$ will not decrease by applying the reduction rules. Let $r=n-|F|$ and $C(r)$ denote the maximum number of leaves in the search tree generated by the algorithm for any instance with measure $r$.
We only need to consider the branching operation in the second step.

For convenience, we call a maximal sequence $P=\{e_1,e_2,\ldots ,e_p\}$ of edges $e_i=u_iu_{i+1}\in E(H)$ $(1\leq i\leq p-1)$  a {\em chain} if all vertices $u_j$ ($j=2,3,\ldots,p-1$) are forced vertices. In the definition of the chain, we allow $u_1=u_p$.
Observe that each chain is contained in the same circuit.
Since the selected circuit ${\cal C}$ has some trivial block, we know that ${\cal C}$ contains at least one chain $P$ of size $\geq 2$.
We distinguish two cases according to the size of $P$ being even or odd.

Case 1. $|P|$ is even: If all the blocks of ${\cal C}$ are trivial, then the $U$-component $H$ containing ${\cal C}$ is a cycle  of even length. Since $H$ cannot be a $4$-cycle, the length of cycle $H$ is at least $6$ (see \reff{ThreeCases}(a) for an illustration).
In each branch, after processing the circuit ${\cal C}$, we can include at least $3$ edges to $F$. This gives us branch vector
 \eqn{basic_1}{  (3;3).}
Next, we assume that  ${\cal C}$ has a nontrivial block. We look at the worst case where $P$ is of size 2 and
 ${\cal C}$ has only one nontrivial block (see \reff{ThreeCases}(b) for an illustration). Now ${\cal C}=P$.
 Let ${\cal C}=\{u_1v_1,u_2v_2\}$, where $v_1=u_2$.
We branch on the circuit at edge $u_1v_1$.
In the branch of deleting $u_1v_1$, we include $u_2v_2$ into $F$ .
Furthermore, we can include the remaining two edges incident on $u_1$ into $F$ by
 simply applying reduction rules. In the other branch of including $u_1v_1$ into $F$, we will delete $u_2v_2$ and include
 the other two edges incident on $v_2$ into $F$. We still can get branch vector \refe{basic_1}.
% \eqn{basic_2}{  C(r)\leq C(r-3) +C(r-3).}
 Note that when ${\cal C}$ is not of the worst case, it is not hard to verify that we may include more edges to $F$ and we can get a branch vector covered by \refe{basic_1}. We omit the details here, since the detailed proof can also be derived from the analysis of the improved algorithm in the next sections.
\vspace{-0mm}\fig{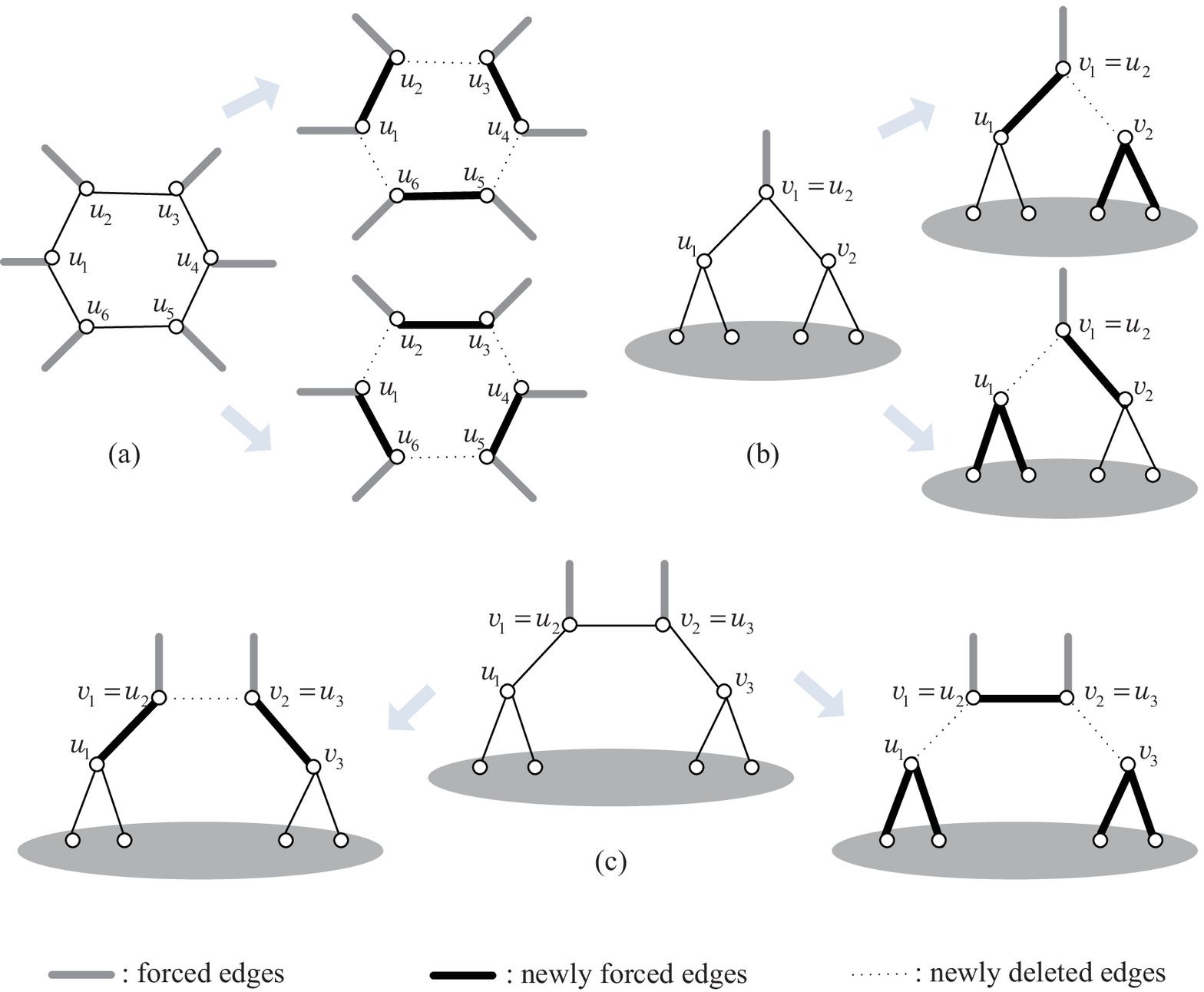}{1}{Three bottleneck cases in branching on a cirucuit ${\cal C}$:
(a) ${\cal C}$ is a 6-cycle; (b) ${\cal C}$ is a chain $P$ of length 2;
(c)  ${\cal C}$ is a chain $P$ of length 3. }{ThreeCases}\vspace{-0mm}

 Case 2. $|P|$ is odd: Now circuit ${\cal C}$ must contain  at least one nontrivial block, otherwise the instance violates the parity condition. We also look at the worst case where $P$ is of size 3 (since $|P|\geq2$) and  ${\cal C}$ has only one nontrivial block (see \reff{ThreeCases}(c) for an illustration). Let ${\cal C}=\{u_1v_1,u_2v_2, u_3v_3\}$, where $v_1=u_2$ and $v_2=u_3$.
We branch on the circuit at edge $u_1v_1$.
In the branch of deleting $u_1v_1$, we will also delete $u_3v_3$  and include the following five edges to $F$: $v_2u_2$, the remaining two edges incident on $u_1$ and the remaining two edges incident on $v_3$ (note that since the graph is reduced, $u_1$ and $v_3$ are not adjacent and then the five edges are different to each other). In the other branch of including $u_1v_1$ to $F$, we will delete $u_2v_2$ and include $u_3v_3$ to $F$. We can get branch vector
 \eqn{basic_3}{  (5;2).}
  When ${\cal C}$ is not of the worst case, we can reduce more edges and get \refe{basic_3} at least.

 Since $C(r)=1.260^r$ satisfies the two recurrences corresponding to \refe{basic_1} and \refe{basic_3}, we know that the simple algorithm can solve the TSP problem in an $n$-vertex degree-3 graph in $O^*(1.260^n)$ time, which achieves the same running time bound of Eppstein's algorithm for TSP3 in~\cite{Eppstein:TSP3}.

\section{The measure and conquer method}\label{sec_MaC}
The measure and conquer method, first introduced by  Fomin, Grandoni and Kratsch~\cite{kn:fomin2}, is one of the most powerful tools to analyze exact algorithms.
It can obtain improved running time bound for many branching-and-search algorithms without making any modification to the algorithms. Currently, many best exact algorithms for NP-hard problems are based on this method.
In the measure and conquer method, we may
set a weight of vertices in the instance and use the sum $w$ of the total weight in the graph as the measure
to evaluate the running time. In the algorithm, the measure $w$ should satisfy the \emph{measure condition}: (i) when $w\leq0$ the instance can be solved in polynomial time;
(ii) the measure $w$ will never increase in each operation in the algorithm; and (iii) the measure will decrease in each of the subinstances generated by applying a branching rule. With these constraints, we may build recurrences for the branching operations.
Next, we introduce a way of applying the measure and conquer method to the above simple algorithm.

The graph has three different {\em vertex-weight}.
For each vertex $v$, we set its vertex-weight $w(v)$
to be
\[ w(v)
=\left\{ \begin{array}{cl}
 w_3=1 & \mbox{if  $d_U(v)=3$ }\\
 w_{3'} & \mbox{if $d_U(v)=2$ and $d_F(v)=1$}\\
 0 & \mbox{otherwise.}   \end{array}
\right.\]
%where we shall determine a best value for $w_{3'}$
%based on our branching rules.
We will determine the best value of $w_{3'}$ such that the worst recurrence in our algorithm is best.
Let $\Delta_3=w_3-w_{3'}$. For a subset of vertices (or a subgraph) $X$, we also use $w(X)$ to denote the total vertex-weight in $X$.

Now we analyze the simple algorithm presented in Section~\ref{simple_alg} by using this vertex-weight setting. Note that since we require that $w_3=1$, the total vertex weight $w$ in the graph is not greater than the number $n$ of vertices in the graph. We can get a running time bound related to $n$ if we get a running time bound related to $w$.
Here we only examine the three bottleneck cases in \reff{ThreeCases}.

When we branch on a circuit of 6-cycle in \reff{ThreeCases}(a),  in  each branch,
all the six forced vertices will be reduced and then we can reduce $w$ by $6w_{3'}$.
We get the following branching vector:
%Let $C(r)$ denote the maximum number of leaves in the search tree generated by the algorithm for any instance with measure $r$
\eqn{mq1}{[6w_{3'}]_2.}
When we branch on a circuit  of chain of length 2 in \reff{ThreeCases}(b), in the branch where $u_1v_1$ is included to $F$, $u_2v_2$ is deleted, and $v_2v'_{2}$ and $v_2v''_{2}$ are also included to $F$ by reduction rules, where $v'_2$ and $v''_2$ are the two neighbors of $v_2$ other than $u_2$. Then we can reduce $w$ by $w_{3'}$ from $v_1$, $\Delta_3$ from $u_1$, $w_3$ from $v_2$, and $2\delta_1$ from $v'_2$ and $v''_2$, where $\delta_1\geq \min\{ \Delta_3,w_{3'} \}$. The second branch can be analyzed in a similar manner. We get
\eqn{mq2}{[w_{3'}+\Delta_3+w_3+2\delta_1]_2=[2+2\delta_1]_2.}
When we branch on a circuit  of chain of length 3 in \reff{ThreeCases}(c), in the branch of including $u_1v_1$ to $F$, we reduce $w$ by $2\Delta_3$ from $u_1$ and $v_3$ and $2w_{3'}$ from $v_1$ and $v_2$. In the other branch, we can reduce $w$ by $2w_3$ from $u_1$ and $v_3$, $2w_{3'}$ from $v_1$ and $v_2$, and $4\delta_1$ from $\{u'_1,u''_1,v'_3,v''_3\}$, where $u'_1$ and $u''_1$ are the two neighbors of $u_1$ other than $v_1$ and $v'_3$ and $v''_3$ are the two neighbors of $v_3$ other than $u_3$. We get
\eqn{mq3}{(2;2w_3+2w_{3'}+4\delta_1).}
We can verify that under that above three constraints, the best value of $w_{3'}$ is ${\frac{1}{2}}w_3=\frac{1}{2}$. With this setting, we can see that \refe{mq1} and \refe{mq2} become \refe{basic_1}, and \refe{mq3} becomes \refe{basic_3}. This also tells us that the measure and conquer method cannot directly derive a better running time bound of the simple algorithm. In the next section, we present a new technique and show an improvement by combining the new technique with the traditional measure and conquer method.

\section{Amortization on connectivity structures}
To improve the time bound by the above simple analysis, we need to use more structural properties of the graph.
Note that for the bottleneck case of \reff{ThreeCases}(a), the above algorithm reduces all the vertices in this $U$-component. It is impossible to improve by reducing more vertices (or edges) and so on. But we also reduce the number of $U$-components by 1. This observation gives us an idea of an amortization scheme over the cut-circuit structure by setting a weight on each $U$-component in the graph.

In this method, each vertex in the graph receives a nonnegative weight as shown in Section~\ref{sec_MaC}. We also set a  weight (which is possibly negative, but bounded from by a constant $c\geq 0$) to each $U$-component. Let $\mu$ be the sum of all vertex weight and $U$-component weight. We will use $\mu$ to measure the size of the search tree generated by our algorithm.  The measure $\mu$ will also satisfy the measure condition. Initially there is only one $U$-component and $\mu< n+c$ holds.
If we get a running time bound related to $\mu$ for our algorithm, then we get a running time bound related to $n$.

A simple idea is to set the same weight to each nontrivial $U$-component.  It is possible to improve the previous best result by using this simple idea. However, to get further improvement, in this paper, we set several different component-weights. Our purpose is to distinguish some ``bad" $U$-components, which will be characterized as  ``critical" $U$-components.

An {\em extension} of a 6-cycle is obtained from a 6-cycle $v_1v_2v_3v_4v_5v_6$ and a 2-clique $ab$
by joining them with two independent edges $av_i$ and $bv_j$ ($i\neq j$).
An extension of a 6-cycle always has exactly eight vertices.
A chord of an extension of a 6-cycle is an edge between two vertices in it but different from the eight edges
$v_1v_2,v_2v_3,v_3v_4,v_4v_5,v_5v_6,v_6v_1,av_i$, $bv_j$ and $ab$.

A subgraph $H$ of a $U$-component in an instance $(G,F)$ is \emph{$k$-pendent critical}, if it is a 6-cycle or an extension of a 6-cycle with
$|\mathrm{cut}_U(H)|=k$ and $|\mathrm{cut}_F(H)|=6-k$
%\mathrm{cut}_F (H)|+|\mathrm{cut}_U(H)|=6$
(i.e., $H$ has no chord of unforced/forced edge). A $0$-pendent critical $U$-component is also simply called a \emph{critical graph} or \emph{critical $U$-component}.
\reff{critical} illustrates two examples of critical graphs of extensions of a $6$-cycle.
Branching on a critical $U$-component may lead to a bottleneck recurrence in our algorithm.
So we set a different component-weight to this kind of components to get improvement.

\vspace{-0mm}\fig{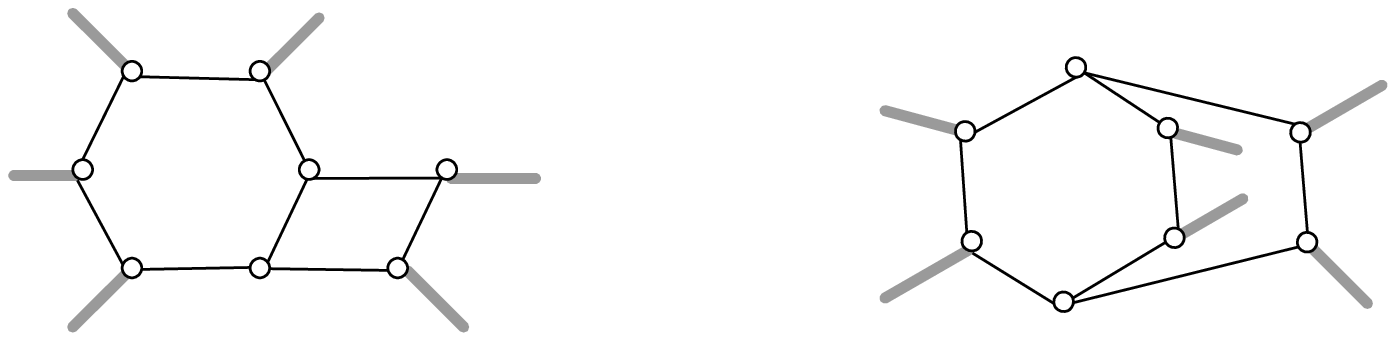}{0.8}{Extensions of a 6-cycle}{critical}\vspace{-0mm}

For each $U$-component $H$, we set its component-weight $c(H)$  to be
\[ c(H)
=\left\{ \begin{array}{cl}
 0 & \mbox{if  $H$ is trivial}\\
 -4w_{3'} & \mbox{if $H$ is a 4-cycle}\\
 \gamma & \mbox{if $H$ is a critical $U$-component}\\
 \delta  & \mbox{otherwise,}   \end{array}
\right.\]
where we set $c(H)=-4w_{3'}$ so that $c(H)+w(H)=0$ holds for
every 4-cycle $U$-component $H$.

We also require that the vertex-weight and component-weight satisfy the following requirements
\eqn{requirment1}{2\Delta_3 \geq \gamma \geq \delta \geq \Delta_3 \geq {1\over 2} w_3,~ w_{3'}\geq {1\over 5}w_3 ~~\mathrm{and}~~\gamma-\delta\leq w_{3'}.}
Under these constraints, we still need to decide the values of $w_{3'}, \gamma$ and $\delta$ such that the time bound
derived by the worst recurrences in our algorithm will be minimized.
In \refe{requirment1}, $2\Delta_3 \geq \gamma$ is important, because it will be used to satisfy the measure condition (ii).
The other constraints in \refe{requirment1} are mainly used to simplify some arguments and they will not become the bottleneck in our analysis.
Next, we first describe our simple algorithm.

\section{The algorithm}%%%%%%%%%%%%%%%%%
%Processing circuit is the only step in the algorithm that may create exponential running time.
%To reduce the running time bound, we choose circuits with some order to branch on. Next we analyze the branch vectors for each possible case.

A block is called a \emph{normal block} if it is none of trivial, reducible and 2-pendent critical.
A normal block is \emph{minimal} if no subgraph of it is a normal block along any circuit.
Note that when $F$ is not empty, each $U$-component has at least one nontrivial circuit.
Our recursive algorithm for forced TSP  only contains two main steps:\\
{\bf 1.} First apply the reduction rules to a given
instance until it becomes a reduced one; and \\
{\bf 2.} Then take any $U$-component $H$ that is neither trivial nor a 4-cycle
(if no such $U$-component $H$, then the instance is polynomially solvable
by \refl{solvable}),
 and branch on a nontrivial circuit ${\cal C}$ in $H$, where
${\cal C}$ is chosen so that \\
(1) no normal block appears along ${\cal C}$
(i.e., ${\cal C}$ has only trivial and 2-pendent critical blocks)  if this kind of circuit exist;
 and \\
(2)  a minimal normal
 block $B_1$ in $H$ appears along ${\cal C}$  otherwise.
%{\tt[XXX: The assumption that `each nontrivial circuit in H has a normal block' is not used now.]}

We  use $\mu$ as the measure to analyze the size of the search tree in our algorithm.
It is easy to see that after applying the reduction rules on a 2-edge-connected instance,
 the resulting instance remains 2-edge-connected. By this observation and \refl{circuitprocess},
we can guarantee that  an input instance is always 2-edge-connected. Our analysis is based on this.

\subsection{Basic properties of the measure}
Before analyzing the time bound on our algorithm,
we first give basic properties of  the measure  $\mu$.

We show that the measure will not increase after applying any reduction operation  in an 2-edge-connected instance.
Since an input instance is 2-edge-connected,  there is no eliminable edge.
In fact, we always deal with eliminable edges in circuit procedures.
For reducible edges, we deal with them during a process of a reducible circuit
(including the reducible edges to $F$ and dealing with the resulting eliminable edges).
We will show that $\mu$ never increases after processing a circuit.
The measure $\mu$ will not increase after deleting any unforced parallel edge.
The following lemma also shows that  applying the $3$/$4$-cut reduction does not increase $\mu$.

\lem{reductionM}{For a given instance, \\
{\rm (i)} applying the $3$-cut reduction does not increase the measure $\mu$; and \\
{\rm (ii)} applying the $4$-cut reduction on a $U$-component $X$ decreases the measure $\mu$  by $w(X)+c(X)$.}
\pf{It is easy to observe (i).
Next we prove (ii). In the case where the resulting graph consists of four single vertices or a pair of forced edges
after applying the $4$-cut reduction,
the whole component $X$ is eliminated, decreasing $\mu$ by $w(X)+c(X)$ and then the lemma holds.
Otherwise the resulting component, say $X'$ is a $4$-cycle,
where $w(X')+c(X')=0$ according to our setting on the component-weight of $4$-cycles,
and $\mu$ again decreases by $w(X)+c(X)$.
 }\bigskip

%By  the setting of the weights and \refl{solvable}, we can see that
%\lem{basic_1}{For a given 2-edge-connected instance $(G,F)$ with an $n$-vertex graph $G$,\\
%(i) the measure $\mu$ is $w(G)+c(G)< n+2$ if $F=\emptyset$; and \\
%(ii) it is polynomially solvable if $\mu\leq0$.
%}
%
%\lem{reduction1}{For a given instance,\\
%(i) the measure $\mu$ will not increase after removing a parallel unforced edge;\\
%(ii) a reducible subgraph can be found in polynomial time; and \\
%(iii) the measure $\mu$ will decrease by $w(X)+c(X)$ after applying the $4$-cycle reduction on a reducible subgraph $X$.
%}
%\pf{It is obvious that (i) and (ii) hold. We only need to consider (iii). }

\bigskip
Next we consider how much amount of measure decreases by processing a circuit.
We consider that the measure  $\mu$ becomes zero whenever
we find an instance infeasible by \refl{infeasibility}.
After processing a circuit
${\cal C}=\{e_i=u_iv_i\mid 1\leq i\leq p\}$ in a $U$-component $H$,
each block $B_i$ along ${\cal C}$ becomes a new $U$-component, which we denote by $\bar {B_i}$.
We define the {\em direct benefit} $\beta'(B_i)$ from $B_i$ to
be the decrease in vertex-weight of the endpoints $v_i$ and $u_{i+1}$ of $B_i$
minus the component-weight $c(\bar {B_i})$ in the new instance after the circuit procedure.
Immediately after the procedure, the measure $\mu$ decreases
by $w(H)+c(H)-\sum_i (w(\bar {B_i})+c(\bar {B_i})) =c(H)+\sum_i\beta'(B_i)$.
%If all edges in $H$  will be
%further processed by reduction procedure, then
%the  benefit  $\beta_H$ of $H$  is defined to be $w(H)$.
After the circuit procedure, we see that
the vertex-weights of endpoints of each non-reducible and nontrivial block $B_i$ decreases
by $\Delta_3$ and $\Delta_3$ (or $w_3$ and $w_3$) respectively if  $B_i$ is even,
and
% while they decreases
 by $\Delta_3$ and $w_3$ (or $w_3$ and $\Delta_3$) respectively if $B_i$ is odd.
%When $\mathrm{cut}_U(B)$ of a critical block $B$ is deleted,
% $B_i$ will be eliminated or replaced with a 4-cycle component
% by \refl{critical}.
%\marginpar{{\tt see PPT6,7}}
Summarizing these, the direct benefit $\beta'(B)$ from a block $B$ is given
by
\begin{equation}\label{beta-1}
 \beta'(B)
=\left\{ \begin{array}{cl}
 0 & \mbox{if  $B$ is reducible},\\
 w_{3'} & \mbox{if  $B$ is trivial},\\
 w_3\!+\!\Delta_3 \!-\!\delta & \mbox{if $B$ is  odd and nontrivial},\\
 \mbox{$2w_3-\delta$}  & \mbox{if $B$ is even and non-reducible, and $\mathrm{cut}_U(B)$ is deleted},\\
 \mbox{$2\Delta_3 -\gamma $}  & \mbox{if $B$ is $2$-pendent critical, and $\mathrm{cut}_U(B)$ is included in $F$},\\
%& \mbox{~~ to $F$}\\
% 2w_3 -\delta  & \mbox{if $B$ is a noncritical even block, and}\\ &\mbox{~~
%$\mathrm{cut}_U(B)$ is (resp., deleted)}\\
\mbox{$w(B)$} & \mbox{if $B$ is a $2$-pendent 4-cycle, and $\mathrm{cut}_U(B)$ is included in $F$}, \\
\mbox{$2\Delta_3 -\delta$} %~(\geq 2w_3+4w_{3'})$)}
& \mbox{otherwise (i.e., $B$ is even, non-reducible but not }\\
& \mbox{~ a $2$-pendent critical $U$-graph or a $2$-pendent 4-cycle,} \\
 &\mbox{~  and $\mathrm{cut}_U(B)$ is included to $F$).}\\
%
%
% w(B)~(\geq 2w_3+4w_{3'}) &
% \mbox{if $B$ is a critical 2-pendent $U$-graph,  and }\\
% &\mbox{~~$\mathrm{cut}_U(B)$ is (resp., deleted)}\\
%\mbox{$2\Delta_3 -\delta $ (resp., $w(B)$)}
% & \mbox{if $B$ is an extension of a critical graph,} \\ &\mbox{~~and
%$\mathrm{cut}_U(B)$ is included to $F$ (resp., deleted),}\\
% w(B)~(\geq 2w_3+6w_{3'}) &
% \mbox{if $B$ is  an extension of a critical graph, } \\
% &\mbox{~~and $\mathrm{cut}_U(B)$ is deleted,}
\end{array}
\right.\end{equation}
%Note that even if $B$ is a critical graph, $B$ is a not a critical $U$-component in the graph after deleting $\mathrm{cut}_U(B)$.

%After processing a circuit ${\cal C}$ in a 2-edge-connected $U$-component $H$, the measure $\mu$ decreases by
%$c(H)+\sum_i \beta'(B_i)$, where $B_i$ are the blocks along circuit ${\cal C}$.
By \refe{requirment1}, we have that $\beta'(B_i)\geq 0$ for any type of block $B_i$, which implies that
the decrease $c(H)+\sum_i \beta'(B_i)\geq c(H)\geq 0$ (where $H$ is not a 4-cycle) is in fact nonnegative,
i.e.,    the measure $\mu$ never increases by processing a circuit.

After processing a circuit ${\cal C}$, a reduction operation may be applicable to some $U$-components $\bar{B}_i$
 and we can decrease $\mu$ more by reducing them. The \emph{indirect benefit} $\beta''(B)$ from a block $B$ is defined as the amount of $\mu$ decreased by applying reduction rules on the $U$-component $\bar{B}$ after processing the circuit. Since we have shown that $\mu$ never increases by applying reduction rules, we know that $\beta''(B)$  is always nonnegative. The \emph{total benefit} (\emph{benefit}, for short) from a block $B$ is
$$\beta(B)=\beta'(B)+\beta''(B).$$

\lem{circuit_reduce1}{After processing a circuit ${\cal C}$ in a $2$-edge-connected $U$-component $H$ $($not necessary being reduced$)$ and applying reduction rules until the instance becomes a reduced one, the measure $\mu$ decreases by
$$c(H)+\sum_i \beta(B_i),$$
where $B_i$ are the blocks along circuit ${\cal C}$.
}\bigskip

The indirect benefit from a block depends on the structure of the block.
In our algorithm, we hope that the indirect benefit is as large as possible.
Here we prove some lower bounds on it for some special cases.

%{\tt[XXX: This lemma is newly added. I donot know the proof OK or not.]}

\lem{further1}{Let $H$ be a $U$-component containing no induced triangle and ${\cal C}'$ be a reducible circuit in it such that there is exactly one reducible block along ${\cal C}'$. The measure $\mu$ decreases by at least $2\Delta_3$ by processing the reducible circuit ${\cal C}'$ and applying reduction rules.
}
\pf{By assumption, the reducible circuit  has at least two blocks, one reducible block $B_1$ and one non-reducible block $B_2$.
(1) Assume that  every other block than $B_1$  along ${\cal C}'$ is trivial.
Then $H$ should be a cycle of length $k\geq 5$ (if $H$ is a 4-cycle, then there are three odd blocks along ${\cal C}'$ and we find the instance infeasible by \refl{infeasibility}).
There are $k-1\geq 4$ trivial blocks along ${\cal C}'$. After processing the circuit,   the whole component $H$ will be eliminated decreasing $\mu$
by at least $c(H)+w(H)\geq \delta+4 w_{3'}\geq 2\Delta_3$ (by \refe{requirment1}).
(2) Otherwise, i.e., there is a block $B_2$ of more than one vertex ($B_2$ is not trivial or reducible):
(2-i)  $\mathrm{cut}_U(B_2)$ is deleted in the circuit procedure:
Then $\mu$ decreases  by at least
$c(H)+\beta'(B_2)=\delta + (2w_3-\delta)=2w_3\geq 2\Delta_3$ by (\ref{beta-1}).
Hence assume that $\mathrm{cut}_U(B_2)$ is included to $F$ in (2).
(2-ii) $B_2$ is not 2-pendent critical:
 Then the measure $\mu$ decreases by at least $c(H)+\beta'(B_2)=\delta + (2\Delta_3-\delta)=2\Delta_3$.
The remaining case is that $B_2$ is 2-pendent critical and $\mathrm{cut}_U(B_2)$ is included to $F$.
(2-iii)  there are only two blocks $B_1$ and $B_2$ along ${\cal C}'$:
By processing  the circuit ${\cal C}'$, the two edges in $\{xz,yz\}=\mathrm{cut}_U(B_2)$ become forced edges.
Since they are incident on the single vertex $z$ in $B_1$,
we can replace $xz$ and $yz$ with a single forced edge $xy$ and then $B_2$ becomes
a 0-pendent 6-cycle or extension of a 6-cycle with only one forced chord $xy$.
After the circuit procedure of ${\cal C}'$, we can apply 4-cut reduction to the 0-pendent 6-cycle $B_2$ (by \refl{reducecritical})
 and then this decreases $\mu$ by $c(H)+\beta(B_2)= \delta+w(B_2)> 2\Delta_3$ (by \refl{reductionM}(ii)).
(2-iv)  there is a pair of  two trivial blocks $B_3$ and $B_4$ or a nontrivial and non-reducible block $B_5$ along ${\cal C}'$:  Now we can decrease $\mu$ by at least $c(H)+\beta(B_2)+\sum_{i\neq 2}\beta(B_i)\geq \delta+(2\Delta_3-\gamma)+\min \{ 2w_{3'}, (2\Delta_3-\delta) \} \geq 2\Delta_3$ (by \refe{requirment1}).

In any case,  the measure $\mu$ decreases by at least $2\Delta_3$.
}

\lem{indirectB}{In the circuit procedure for a circuit ${\cal C}$ in a reduced instance, the indirect benefit from a block $B$ along ${\cal C}$ satisfies
\[ \beta''(B)
\geq \left\{ \begin{array}{clc}
 2\Delta_3 & \mbox{if $B$ is odd and nontrivial}, & {\rm (i)}\\
% 0 & \mbox{if $B$ is a $2$-pendent 4-cycle},&\\
% &  \mbox{~~and $\mathrm{cut}_U(B)$ is included to $F$},& {\rm (ii)}\\
 w(B)-\beta'(B) & \mbox{if $B$ is a $2$-pendent cycle or critical graph},&\\
 & \mbox{~~and $\mathrm{cut}_U(B)$ is deleted},& {\rm (ii)}\\
 \delta  & \mbox{if $B$ is even but not reducible or a $2$-pendent}&\\
 & \mbox{~~ cycle, and $\mathrm{cut}_U(B)$ is deleted},& {\rm (iii)}\\
 0 & \mbox{otherwise}.& {\rm (iv)}\\
\end{array}
\right.\]
}

\pf{We will use $\bar B$ to denote the $U$-component resulting from $B$ after the circuit procession.
Case (i): Since $B$ is odd,
only one edge in $\mathrm{cut}_U(B)$ is deleted (the other one is included to $F$) in the circuit procession. Then there is exactly one vertex of degree 2 in $\bar B$, which is the only reducible block along a circuit ${\cal C}'$ in $\bar B$. Since the original instance is reduced and contains no triangle, we know that circuit ${\cal C}'$ satisfies the condition in \refl{further1}. By processing ${\cal C}'$ in $\bar B$, we can decrease $\mu$ by at least $2\Delta_3$ by \refl{further1}.
Then we get $\beta''(B)\geq 2\Delta_3$.
%
%Case (ii) holds because after the circuit procedure $\bar B$ becomes a $U$-component of a $4$-cycle and we will not apply any other operations on it any more.
%
For case (ii),  if $B$ is a $2$-pendent cycle, we can reduce the whole $U$-component $\bar B$ after the circuit
procedure, and then we have $\beta''(B)=c(\bar B)+w(\bar B)$.
Otherwise  $B$ in (ii) is a $2$-pendent critical graph and the 4-cut reduction can be applied to $\bar B$
 by  \refl{reducecritical} since the two end vertices of edges in $\mathrm{cut}_U(B)$ will be of degree 2 in  $\bar B$
after $\mathrm{cut}_U(B)$ is deleted.
Then we still have $\beta''(B)=c(\bar B)+w(\bar B)$ by \refl{reductionM}(ii). Note that $\beta'(B)= w(B)-w(\bar B)-c(\bar B)$ (by the definition of $\beta'(B)$). We get
$\beta''(B)=w(B)-\beta'(B)$.
%{\tt [NNN: not clear to me how $w(B)-\beta'(B)$ comes from $w(X)+c(X)$ of \refl{reductionM}(ii)][XXX: I have added more explanations. Is it clear now?]}
%
For Case (iii), we will get at least one reducible circuit ${\cal C}'$ in $\bar B$. Note that $\bar B$ has some vertex of degree 2 and cannot be a critical graph.
Hence $c(\bar B)=\delta$.
By processing the reducible circuit ${\cal C}'$, we can decrease $\mu$
 by at least $c(\bar B)+\sum_i\beta(B'_i)\geq c(\bar B)=\delta$, where $B'_i$ are the blocks along ${\cal C}'$.
 The inequality in (iv) holds since we have proved that $\mu$ will never increase after applying reduction rules.
 }

\section{The Analysis}%%%%%%%%%%%%%%%%%%%
Now we are ready to analyze our algorithm.
In the algorithm,   branching on a circuit  generates two instances $(G_1,F_1)$ and $(G_2,F_2)$.
By \refl{circuit_reduce1}, we get branch vector
$$(c(H)+\sum_i \beta_1(B_i); c(H)+\sum_i \beta_2(B_i)),
$$
where  $\beta_j(B)$, $\beta'_j(B)$ and $\beta''_j(B)$  denote
the functions $\beta(B)$, $\beta'(B)$ and $\beta''(B)$ evaluated in $(G_j,F_j)$, $j=1,2$
for clarifying how branch vectors are derived in the subsequent analysis.
We consider this branch vector for different cases.
First we analyze the easy case where the chosen circuit ${\cal C}$ has no normal block.
Then  we analyze the somewhat complicated case
where there is a minimal normal block along ${\cal C}$.

\subsection{Circuits with only trivial and $2$-pendent critical blocks}
In this subsection, we assume that the chosen circuit ${\cal C}$ in a $U$-component $H$ (not a 4-cycle)
has only trivial and $2$-pendent critical blocks.
% a reduced instance has  a $U$-component $H$ (not a 4-cycle)
%in which no normal block appears appear along any circuit
%Let ${\cal C}$ be an arbitrary circuit in $H$,
%where every block along ${\cal C}$ is  trivial or a $2$-pendent critical block.
We consider the following three cases.

{\bf Case 1.} All blocks along ${\cal C}$ are trivial blocks: Now $H$ should be a cycle of even length $l\geq 6$.
By (\ref{beta-1}) and \refl{circuit_reduce1}, we know that
in each branch we can decrease $\mu$ by at least
 \[ c(H)+\sum_i \beta(B_i)
\geq \left\{ \begin{array}{cl}
 \gamma+ 6 w_{3'}& \mbox{if  $l=6$ }\\
 \delta +8 w_{3'} & \mbox{if $l\geq 8$.}   \end{array}
\right.\]
Then we get branch vectors
 \eqn{6-cycle}{ % [ c(H)+w(H)]_2=
[6w_{3'}+\gamma]_2 }
 % c(1)= -1+2*x(2)^(-(6*x(1) +x(3) )); %QP{6-cycle}
for  $l=6$
and %$[c(H)+8w_{3'}]_2=
$[ 8w_{3'}+\delta]_2$ for $l\geq 8$,
 which is covered by \refe{6-cycle}.
Note that \refe{6-cycle} is tight for a circuit of 6-cycle in \reff{ThreeCases}(a).

{\bf Case 2.} There is only one 2-pendent critical block $B_1$ along ${\cal C}$:
Since $B_1$ is even, the number $r>0$ of trivial blocks (odd blocks) along ${\cal C}$ is also even by the parity condition.
%Case 2.1. $B_1$ is a 2-pendent extension of a 6-cycle. Then $w(B_1)=4w_3+4w_{3'}$ and $c(H)=\delta$.
%In the branch where $\mathrm{cut}_U(B_1)$ is deleted, $\mu$ decreases by $c(H)+\sum_i \beta(B_i)\geq \delta+w(B_1)+r w_{3'}\geq \delta +4w_3+6w_{3'}$. In the branch where $\mathrm{cut}_U(B_1)$ is included to $F$, $\mu$ decreases by $c(H)+\sum_i \beta(B_i)\geq \delta+r w_{3'}\geq \delta +2w_{3'}$. This gives branch vector
% \eqn{6-cycle-extension1}{
%(\delta+4w_3+6w_{3'}; \delta +2w_{3'}).
% }
% c(2')=-1+ x(2)^(-(4+6*x(1)+x(4)))+x(2)^(-(2*x(1)+x(4))); %QP{extension of 6-cycle-1}
%
%
%We show that $B_1$ can only be a 2-pendent 6-cycle.
%To show this, assume that $B_1$ is a 2-pendent extension of a 6-cycle.
%
%I this case there is a pair of adjacent forced vertices in $B_1$ (since $B_1$ has four forced vertices) and
%three unforced edges incident on them would form a circuit with a normal block and two trivial blocks
%{\tt [NNN: why no other circuit (say of length 5)?]},
%contradicting the assumption on $H$.
%This proves that $B_1$ is always a 2-pendent 6-cycle, and we have $\beta(B_1)=2w_3+4 w_{3'}$
%{\tt [NNN: how this is obtained? Use some lemmas?]}.
%Since $B_1$ is even, the number $r>0$ of trivial blocks (odd blocks) along ${\cal C}$ is also even by the parity condition.
Note that if $r=2$ and $B_1$ is 2-pendent 6-cycle then $H$ is a critical graph (an extension of a 6-cycle) and $c(H)=\gamma$. Otherwise $H$ is not critical and $c(H)=\delta$.
Then in the branch where $\mathrm{cut}_U(B_1)$ is deleted, the  decrease $c(H)+\sum_i \beta(B_i)$ of $\mu$ is at least
$c(H)+w(B_2)+r w_{3'}
\geq \min\{ \gamma+ (2w_3+4w_{3'})+2w_{3'}, \delta + (4w_3+ 4w_{3'})+2w_{3'}, \delta +  (2w_3+4 w_{3'})+4 w_{3'} \}
=$
\[\gamma+ 2w_3+6 w_{3'}.
\]
In the branch where $\mathrm{cut}_U(B_1)$ is included to $F$, the  decrease $c(H)+\sum_i \beta(B_i)$ of $\mu$ is at least
\[\min\{ \gamma, \delta\}+ (2\Delta_{3}-\gamma)+2 w_{3'}=\delta+2w_3-\gamma.
\]
This gives branch vector
 \eqn{6-cycle-extension}{
(\gamma+2w_3+6w_{3'}; \delta+2w_3-\gamma).
 }
% c(2)= -1+ x(2)^(-(2+6*x(1)+1*x(3)))+x(2)^(-(2-x(3)+x(4))); %QP{extension of 6-cycle}
% {\tt[XXX: Case 2 is modified according to the algorithm. %Some arguments are simplified but we should relax more. I updated \refe{6-cycle-extension}. The finial result is not changed. But the domain of $\delta$ is smaller. ]}

{\bf Case 3.} There are at least two 2-pendent critical blocks $B_1$ and $B_2$ along ${\cal C}$:
If $\mathrm{cut}_U(B_2)$ is included to $F$ in the branch where $\mathrm{cut}_U(B_1)$ is deleted,
then we can get branch vector
$ [c(H)+\beta(B_1)+\beta(B_2)]_2=[\delta+(2w_3+4w_{3'})+(2\Delta_{3}-\gamma)]_2=[\delta+4w_3+2w_{3'}-\gamma]_2$, which is covered by \refe{6-cycle}.
On the other hand, if $\mathrm{cut}_U(B_2)$ is also deleted in the branch where $\mathrm{cut}_U(B_1)$ is deleted, we get branch vector
\eqn{good2a}{(\delta+2(2\Delta_3-\gamma); \delta+2(2w_3+4w_{3'})).}
% c(3)= -1+x(2)^(-(4-4*x(1) -2*x(3)+x(4)))+x(2)^(-(4+8*x(1)+x(4))); %QP{good2a} %QP{2 6-cycles}

\subsection{Circuits with a minimal normal block}
In this subsection, we assume that the chosen circuit ${\cal C}$ in a $U$-component $H$
has a minimal normal block $B_1$. Now $c(H)=\delta$ always holds.
%In this subsection, we can assume that a reduced instance has a $U$-component $H$ that contains
% a circuit ${\cal C}$ along which there is a minimal normal block $B_1$.
%{\tt [$c(H)=\delta$ should be stated here?]}
%Our algorithm will branch on  circuit ${\cal C}$.
We distinguish several cases to analyze the branch vectors.

{\bf Case 1.} Block $B_1$ is odd: Circuit  ${\cal C}$ has another odd block $B_2$.
By (\ref{beta-1}) we have  $\beta(B_2)\geq \beta'(B_2)=\min\{w_{3'}, w_3\!+\!\Delta_3 \!-\!\delta\}= w_{3'}$
  in each branch.
For $B_1$, we have that
$\beta'(B_1)=w_3+\Delta_3 -\delta$ and $\beta''(B_1)\geq 2\Delta_3$ by \refl{indirectB}.
In each branch,  $\mu$  decreases by at least
$c(H)+\sum_i \beta(B_i)\geq \delta+ \beta'(B_1)+ \beta''(B_1)+\beta(B_2)\geq \delta + (w_3+\Delta_3 -\delta)+ 2\Delta_3+w_{3'}=4w_3-2w_{3'}$.
Therefore, we can get
branch vector
\eqn{even1}{  [4w_3-2w_{3'}]_2.}
%c(4)= -1+2*x(2)^(-(4-2*x(1) )); %QP{even length}
Note that \refe{even1} is tight for a circuit of chain of length 2 in \reff{ThreeCases}(b).

\bigskip
Next, we assume that $B_1$ is even.
In the branch where $\mathrm{cut}_U(B_1)$ is included to $F$, we have that $\beta'_1(B_1)=2\Delta_3-\delta$.
In the other branch, we have that $\beta'_2(B_1)=2w_3-\delta$.
%Then we get
%$$ [c(H)+\sum_i \beta_{B_i}]$$
By branching on ${\cal C}$, we get branch vector
%$$ (c(H)+\sum_i \beta_1(B_i), c(H)+\sum_i \beta_2(B_i) )=$$
$$ (\delta+\beta'_1(B_1)+\beta''_1(B_1)+\xi_1; \delta+\beta'_2(B_1)+\beta''_2(B_1)+\xi_2),
$$
where $\xi_j=\sum_{i\neq1} \beta_j(B_i)\geq 0$ ($j\in \{1,2\}$), $\beta'_1(B_1)=2\Delta_3-\delta$ and $\beta'_2(B_1)=2w_3-\delta$.
We here evaluate $\xi_1$ and $\xi_2$.
If all blocks other than $B_1$ along ${\cal C}$ are 2-pendent critical, then we have that $\xi_1\geq 2\Delta_3-\gamma$ and $\xi_2=\sum_{i\neq 1}w(B_i)\geq w(B_2)\geq 2w_3+4w_{3'}$.
Otherwise, $\xi_1\geq \min \{2w_{3'}, w_3+\Delta_3-\delta, 2\Delta_3-\delta \}
=2w_{3'}$ and $\xi_2\geq \min \{2w_{3'}, w_3+\Delta_3-\delta, 2w_3-\delta \}=2w_{3'}$.
We have the following two choices for $(\xi_1 , \xi_2)$:
\eqn{xi_bound}{
(\xi_1 , \xi_2) = (2\Delta_3-\gamma , 2w_3+4w_{3'}) \ \ \  {\rm and} \ \ \ (2w_{3'} , 2w_{3'}).
}
%Sometimes we can relax more by simply considering
%\eqn{choice_2}{(\beta_1, \beta_2) = (\min \{2\Delta_3-\gamma, 2w_{3'}\}, \min\{\gamma+ 2w_3+4w_{3'}, 2w_{3'}\})=(2\Delta_3-\gamma,2w_{3'}).}
In what follows,   we derive some lower bounds on $\beta''_1(B_1)$ and $\beta''_2(B_1)$
by examining the structure of $B_1$.

 Let  $\mathrm{cut}_U(B_1)=\{xv,yu\}$,
where $x$ and $y$ are in $B_1$. Let $(G_1, F_1)$ and $(G_2,F_2)$ be the two resulting instances after branching and processing the circuit ${\cal C}$, where $(G_1, F_1)$ corresponds to the branch where $\mathrm{cut}_U(B_1)$ is included to $F$ and $(G_2, F_2)$ corresponds to the branch where $\mathrm{cut}_U(B_1)$ is deleted. Let $x_1x$ and $x_2x$ (resp., $y_1y$ and $y_2y$) be the two unforced edges incident on $x$ (resp., $y$) in $(G_1,F_1)$. Note that $x_1x$ and $x_2x$ (resp., $y_1y$ and $y_2y$) will be in the same circuit ${\cal C}_x$ (resp., ${\cal C}_y$) in $(G_1,F_1)$, since $x$ (resp., $y$) is a forced vertex now.
See \reff{normalblock} for illustrations of the structure of the edges incident to $x$ and $y$.
\vspace{-0mm}\fig{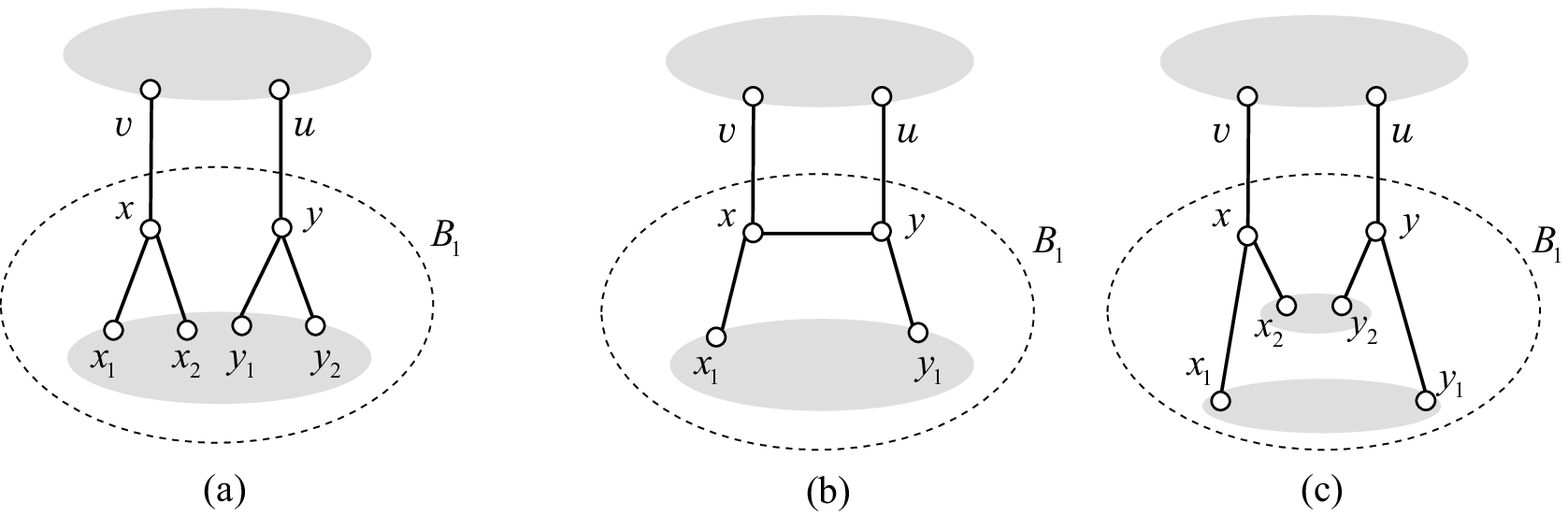}{1}{Illustrations of block $B_1$: (a) ${\cal C}_x\neq{\cal C}_y$;
(b) ${\cal C}_x={\cal C}_y$ and $x$ and $y$ are adjacent;
(c) ${\cal C}_x={\cal C}_y$ and $x$ and $y$ are not adjacent.}{normalblock}\vspace{-0mm}

%{\tt[XXX: There is no (c) in \reff{normalblock}. Could you modify it? I cannot edit the figure now.]}

{\bf Case 2.} Block $B_1$ is even and ${\cal C}_x$ and ${\cal C}_y$ are two different circuits in $(G_1,F_1)$
(see \reff{normalblock}(a)):
Now ${\cal C}_x$ and ${\cal C}_y$ are two different circuits also  in $(G_2,F_2)$.
In the branch where $\mathrm{cut}_U(B_1)$ is deleted, we have
$\beta''_2(B_1)\geq 4\Delta_3$  (by applying \refl{further1} to ${\cal C}_x$ and ${\cal C}_y$).
%Then we get
%$$ [c(H)+\sum_i \beta_{B_i}]$$
Next we consider the other blocks along ${\cal C}$.

Case 2.1. There are at least two odd blocks $B_2$ and $B_3$ along ${\cal C}$:
By (\ref{beta-1}), $\beta_1(B_i)\geq w_{3'}$ and $\beta_2(B_i)\geq w_{3'}$ ($i\in\{2,3\}$).
Since $c(H)+\sum_i \beta(B_i)\geq \delta+ \beta(B_1)+\beta(B_2)+\beta(B_3)$, we get branch vector
\eqn{odd_chain3}{ (\delta+(2\Delta_3-\delta)+2w_{3'};\delta+(2w_3-\delta)+4\Delta_3+2w_{3'})=(2w_3; 6w_3-2w_{3'}).}
% c( )= -1+x(2)^(-(2+0*x(1)))+x(2)^(-(6-2*x(1))); %QP{odd_chain3} %QP{case 2.1}

Case 2.2. There is no odd block along ${\cal C}$:
 Let $B_2$ be the block along ${\cal C}$ containing vertex $v$.
Then $B_2$ is an even block
such that $\mathrm{cut}_U(B_2)$ is also included to $F$ in the branch
 where $\mathrm{cut}_U(B_1)$ is included to $F$.
Then $c(H)+\sum_i \beta(B_i)\geq \delta+ \beta(B_1)+\beta(B_2)$.
If $B_2$ is not a 2-pendent critical block, then $\beta'_1(B_2)= 2\Delta_3-\delta$, $\beta'_2(B_2)=2w_3-\delta$ and $\beta''_2(B_2)\geq \delta$ (by \refl{indirectB}), and we get branch vector
%$$(\delta+2(2\Delta_3-\delta), \delta+(2w_3-\delta)+ 4\Delta_3+(2w_3-\delta)+\delta)=$$
\eqn{new_1}{ (\delta+2(2\Delta_3-\delta); \delta+(2w_3-\delta)+ 4\Delta_3+2w_3)=(4\Delta_3-\delta; 4w_3+4\Delta_3).
}
% c(6)= -1+ x(2)^(-(4-4*x(2)-x(4)))+ x(2)^(-(8-4*x(1))); %QP{new 1}
Otherwise $B_2$ is a 2-pendent critical block. Then $\beta'_1(B_2)= 2\Delta_3-\gamma$
and $\beta_2(B_2)=\beta'_2(B_2)+\beta''_2(B_2)=w(B_2)\geq 2w_3+4w_{3'}$ (by (\ref{beta-1}) and \refl{indirectB}).
We get branch vector
$(\delta+(2\Delta_3-\delta)+(2\Delta_3-\gamma); \delta+ (2w_3-\delta)+ 4\Delta_3+(2w_3+4w_{3'}))$, i.e.,
\eqn{new_2}{ (4\Delta_3-\gamma; 8w_3).
}
%c(7)= -1+ x(2)^(-(4-4*x(2)-x(3)))+ x(2)^(-8); %QP{new 2}

{\bf Case 3.} Block $B_1$ is even and ${\cal C}_x={\cal C}_y$ in $(G_1,F_1)$ (see \reff{normalblock}(b),(c)):
 First of all, we consider $\beta''_1(B_1)$, $\beta''_2(B_1)$ and others.
%The values of $\xi_1$ and $\xi_2$ will be discussed later.
  We look at the circuit ${\cal C}_x$ in $(G_1,F_1)$.
 Except blocks $\{x\}$ and $\{y\}$, there are some other blocks along ${\cal C}_x$.
Note that each block $B'$ along ${\cal C}_x$ should be a trivial or 2-pendent critical block since $B_1$ is a minimal normal block.
We distinguish three cases by considering the number of critical blocks along ${\cal C}_x$.

Case 3.1. $B_1$ is a 2-pendent cycle of length $\ell$ (all blocks along ${\cal C}_x$ are trivial in $(G_1,F_1)$):
Then $\ell$ is an even integer with $\ell=4$ or $\ell\geq 8$   (since $B_1$ is even and non-critical). When $\ell=4$, in both branches, we have
$\beta(B_1)=w(B_1)=2w_3+2w_{3'}$.
Then we can branch with a branch vector $[c(H)+\beta(B_1)]_2=[\delta+2w_3+2w_{3'}]_2$  covered by \refe{6-cycle}.
Next we assume that $\ell\geq 8$.
%In the branch where $\mathrm{cut}_U(B_1)$ is  included to $F$, we have $\beta_1(B_1)=2\Delta_3-\delta$ and
%$\sum_{i\neq 1}\beta_1(B_i)\geq \min \{2w_{3'}, 2\Delta_3-\delta,  2\Delta_3-\gamma\}=2\Delta_3-\gamma$.
%In the other branch, we have $\beta_2(B_1)=w(B_1)=(\ell-2)w_{3'}+2w_3$ and $\sum_{i\neq 1}\beta_2(B_i)\geq 2w_{3'}$.
Now we may  get only $\beta'_1(B_1)=2\Delta_3-\delta$ and $\beta''_1(B_1)\geq 0$ instead of $\beta_1(B_1)=w(B_1)$.
 But it holds that $\beta_2(B_1)=w(B_1)=2w_3+(\ell-2)w_{3'}\geq 2w_3+6 w_{3'}$.
%Let $\ell=8$.
We get branch vector
\eqn{new_m1}{(\delta+(2\Delta_3-\delta)+\xi_1; \delta+(2w_3+6 w_{3'})+\xi_2).}

%This branch vector may not be good enough in our analysis. Note that in $(G_1,F_1)$, $B_1$ becomes an $\ell$-cycle and
%we can further branch on the circuit in $B_1$ with
%$$ [\delta+\ell w_{3'}]_2.$$
%By combining the above two branches together and letting $\ell=8$, we get branch vector
%\eqn{new_3}{ ([\delta+4w_3+4w_{3'}-\gamma]_2, \delta+2w_3+8 w_{3'}).}

Case 3.2. There are only three blocks $\{x\}$, $\{y\}$ and $B'$ along ${\cal C}_x$ in  $(G_1,F_1)$,
where $B'$ is a 2-pendent critical block:
Now $x$ and $y$ are adjacent (see \reff{normalblock}(b)). We assume that $x_2=y$, $y_2=x$, and $x_1,y_1\in B_1$.
We look at $(G_2,F_2)$ wherein $x$ and $y$ are degree-2 vertices.
After including edges $xy$, $xx_1$ and $yy_1$ to $F$,
$B'$ becomes a $4$-cut reducible graph by \refl{reducecritical}.
Then $\mu$ decreases by $w(B')$
after applying the 4-cut reduction. Then we know that $\beta_2(B_1)=w(B_1)=w(x)+w(y)+w(B')\geq 4w_3+4w_{3'}$.
Therefore, we get branch vector
\[ (\delta+(2\Delta_3-\delta)+\xi_1; \delta+(4w_3+4w_{3'})+\xi_2),\]
which is covered by \refe{new_m1}.

Case 3.3. $B_1$ is not a 2-pendent cycle and there are more than three blocks along ${\cal C}_x$ in  $(G_1,F_1)$:
Then there is a nontrivial and nonreducible block $B'_1$ along ${\cal C}_x$ in  $(G_1,F_1)$, where
 $B'_1$ is a 2-pendent critical block since $B_1$ is a minimal normal block.
%For this case, we may just get $\beta''_1(B_1)\geq0$ and $\beta''_2(B_1)\geq 4\Delta_3+\delta-2\gamma$ (by \refl{further2}).
For this case, we only get the following branch vector by branching on ${\cal C}$:
$(\delta+(2\Delta_3-\delta)+\xi_1; \delta+(2w_3-\delta)+\beta''_2(B_1)+\xi_2)=$
\eqn{basic1}{ (2\Delta_3+\xi_1; 2w_3+\beta''_2(B_1)+\xi_2).
}
In fact, the branch vector \refe{basic1} in Case~3.3 can be the bottleneck in the analysis of our algorithm.
However, in $(G_1,F_1)$, circuit ${\cal C}_x$ is a circuit with only trivial and 2-pendent critical blocks.
In our algorithm, circuit ${\cal C}_x$ will be one of the circuits for the next branching,
and it will never be destroyed until we branch on it.
% (no reduction operations or other branching operations will be applied to  the $U$-graph containing ${\cal C}_x$ {???}).
%For the purpose of analysis, we assume that the algorithm branches on ${\cal C}_x$ in $(G_1,F_1)$ after branching on ${\cal C}$
In fact,  branching on ${\cal C}_x$ proves a  branch vector better than \refe{basic1}.
For the purpose of analysis, we derive a branch vector for the three branches, i.e.,
 the branching on ${\cal C}$ followed by  the branching on
${\cal C}_x$ in $(G_1,F_1)$ in the branch of including $\mathrm{cut}_U(B_1)$  into $F$.
%to generate two instances $(G_{1,p},F_{1,p})$, $p=1,2$.
%We derive a branch vector for the three branches $(G_{1,1},F_{1,1})$, $(G_{1,2},F_{1,2})$ and $(G_2,F_2)$.

In Case~3.3, we see that:  either
($a$) ${\cal C}_x$ has at least two trivial blocks $B'_2$ and $B'_3$ different from $\{x\}$ and $\{y\}$ (since the number of odd blocks is even); or ($b$) all blocks other than $\{x\}$ and $\{y\}$  are 2-pendent critical.

Case ($a$): There is also a 2-pendent critical block $B'_1$ along ${\cal C}_x$ (since $B_1$ is not a 2-pendent cycle). In $(G_2,F_2)$, we can see that $\beta''_2(B_1)\geq c(B_1)+w(B'_2)+w(B'_3)=\delta+2w_{3'}$ (since $B'_2$ and $B'_3$ are trivial blocks). In  $(G_1,F_1)$, by branching on ${\cal C}_x$, we can get branch vector $ (c(B_1)+\beta_1(\{x\})+\beta_1(\{y\})+\sum_{i=1}^3\beta_1(B'_i);
c(B_1)+\beta_2(\{x\})+\beta_2(\{y\})+\sum_{i=1}^3\beta_2(B'_i))=
(\delta+4w_{3'}+(2\Delta_3-\gamma); \delta+4w_{3'}+(2w_3+4w_{3'}))=$
%{\tt [NNN: $B'_{\{x\}}$ and $B'_{\{y\}}$ are defined somewhere?]}
\[
(\delta+2w_3+2w_{3'}-\gamma; \delta+2w_3+8w_{3'}).
\]
By combining it with \refe{basic1} and taking $\beta''_2(B_1)=\delta+2w_{3'}$, we get branch vector
\eqn{new_m2}{
(2\Delta_3 \!+\! \xi_1 \!\!+\! (\delta \!+\! 2w_3 \!+\! 2w_{3'} \!-\! \gamma);
2\Delta_3 \!+\! \xi_1 \!\!+\! (\delta \!+\! 2w_3 \!+\! 8w_{3'}\!);
2w_3 \!+\! \delta \!+\! 2w_{3'} \!+\! \xi_2).~~~~
}

Case ($b$): There are at least two 2-pendent critical blocks $B'_1$ and $B'_2$ along ${\cal C}_x$ (since there are at least four blocks among ${\cal C}_x$).
In $(G_2,F_2)$, we may  get only $\beta''_2(B_1)\geq \delta$.
In  $(G_1,F_1)$, by branching on ${\cal C}_x$,  at least we can get branch vector
$$(\delta+2w_{3'}+2(2\Delta_3-\gamma); \delta+2w_{3'}+2(2w_3+4w_{3'}).$$
%\[ (\delta+4w_3-2w_{3'}-2\gamma; \delta+4w_3+10w_{3'}). \]
By combining it with \refe{basic1} and taking $\beta''_2(B_1)=\delta$, we get branch vector
\eqn{new_m3}{
(2\Delta_3 \!+\! \xi_1 \!+\! (\delta \!+\! 4w_3 \!-\! 2w_{3'} \!-\! 2\gamma);
2\Delta_3 \!+\! \xi_1 \!+\! (\delta \!+\! 4w_3 \!+\! 10w_{3'});
2w_3 \!+\! \delta \!+\! \xi_2).
}

Finally, by replacing $\xi_1$ and $\xi_2$  in \refe{new_m1}, \refe{new_m2} and \refe{new_m3} respectively
with the bounds in \refe{xi_bound}, we get the following six branch vectors
\eqn{new_3_1}{
%(2\Delta_3+(2\Delta_3-\gamma); \delta+(2w_3+6 w_{3'})+(2w_3+4w_{3'}))=
(4\Delta_3-\gamma; \delta+4w_3+10 w_{3'}),}
\eqn{new_3_2}{
%(2\Delta_3+2w_{3'}; \delta+(2w_3+6 w_{3'})+2w_{3'})=
(2w_3; \delta+2w_3+8w_{3'}),
}
\eqn{new_3_3}{
(\delta+6w_3-2w_{3'}-2\gamma; \delta+6w_3+4w_{3'}-\gamma;
\delta+4w_3+6w_{3'}),
}
\eqn{new_3_4}{
(\delta+4w_3+2w_{3'}-\gamma;\delta+4w_3+8w_{3'};
\delta+2w_3+4w_{3'}),
}
\eqn{new_3_5}{
(\delta+8w_3-6w_{3'}-3\gamma;\delta+8w_3+6w_{3'}-\gamma;
\delta+4w_3+4w_{3'}),
}
and
\eqn{new_3_6}{
(\delta+6w_3-2w_{3'}-2\gamma;\delta+6w_3+10w_{3'};
\delta+2w_3+3w_{3'}).
}

%c(8)= -1+ x(2)^(-(4-4*x(1)-x(3)))+ x(2)^(-(4+10*x(1)+x(4))); %QP{new 3-1}
%c(9)= -1+ x(2)^(-2)+ x(2)^(-(2+8*x(1)+x(4))); %QP{new 3-2}
%c(10)= -1+ x(2)^(-(6-2*x(1)-2*x(3)+x(4)))+ x(2)^(-(6+4*x(1)-x(3)+x(4)))+ x(2)^(-(4+6*x(1)-0*x(3)+x(4))); %QP{new 3-3}
%c(11)= -1+ x(2)^(-(4+2*x(1)-x(3)+x(4)))+ x(2)^(-(4+8*x(1)-0*x(3)+x(4)))+ x(2)^(-(2+4*x(1)-0*x(3)+x(4))); %QP{new 3-4}
%c(12)= -1+ x(2)^(-(8-6*x(1)-3*x(3)+x(4)))+ x(2)^(-(8+6*x(1)-1*x(3)+x(4)))+ x(2)^(-(4+4*x(1)-0*x(3)+x(4))); %QP{new 3-5}
%c(13)= -1+ x(2)^(-(6-2*x(1)-2*x(3)+x(4)))+ x(2)^(-(6+10*x(1)-0*x(3)+x(4)))+ x(2)^(-(2+3*x(1)-0*x(3)+x(4))); %QP{new 3-6}

\subsection{Overall analysis}

A quasiconvex program is obtained from \refe{requirment1} and 13 branch vectors
(from \refe{6-cycle} to \refe{new_2} and from \refe{new_3_1} to \refe{new_3_6})
%(\refe{6-cycle}, \refe{6-cycle-extension}, \refe{good2a}, \refe{even1}, \refe{odd_chain3}, \refe{new_1}, \refe{new_2},\refe{new_3_1},\refe{new_3_2},\refe{new_3_3},\refe{new_3_4},\refe{new_3_5} and \refe{new_3_6})
in our analysis.
There is a general method to solve quasiconvex programs \cite{eQP}.
For our quasiconvex program, we observe a simple way to solve it.
We look at \refe{6-cycle} and \refe{even1}. Note that $\min \{6w_{3'}+\gamma, 4w_3-2w_{3'}\}$ under the constraint $2\Delta_3 \geq \gamma$ gets the maximum value at the time when $6w_{3'}+\gamma=4w_3-2w_{3'}$ and $2\Delta_3 =\gamma$. We get $w_{3'}={1 \over 3}$ and $\gamma={4\over 3}$. With this setting, we can verify that when $\delta\in [1.2584, 1.2832]$, all branch vectors other than
\refe{6-cycle} and \refe{even1} in our quasiconvex program will not be the bottleneck.
We get a time bound $O^*(\alpha^{\mu})$ with $\alpha=2^{3\over 10}<1.2312$ by setting
$w_{3'}={1 \over 3}$,$\gamma={4\over 3}$ and $\delta\in [1.2584, 1.2832]$ for our problem.
The bottlenecks in the analysis are   \refe{6-cycle}, \refe{even1} and $2\Delta_3 \geq \gamma$ in \refe{requirment1}.

\thm{result}{TSP in an $n$-vertex graph $G$ with maximum degree 3
can be solved  in $O^*(1.2312^n)$ time and polynomial space.}

\section{Concluding Remarks}\label{sec:conclude}

In this paper, we have presented an improved exact algorithm for TSP
in  degree-3 graphs.
The basic operation in the algorithm is to process the edges
in a circuit by either
including an edge in the circuit to the solution or excluding
 it from the solution.
The algorithm is analyzed by using the measure and conquer method and
an amortization scheme over the cut-circuit structure of graphs,
wherein we introduce not only weights of vertices but also
weights of $U$-components to define
the measure of an instance.

The idea of amortization schemes introducing weights on
components may yield better bounds for other exact algorithms
for graph problems if
how reduction/branching procedures change the system of
components  is successfully analyzed.

\end{document}